\documentclass[aps,prd,superscriptaddress,nofootinbib,tighten,preprint]{revtex4}
\usepackage{color}
\usepackage{latexsym}
\usepackage{amssymb}
\usepackage{amsmath}
\usepackage{graphicx}
\usepackage{hyperref}
\usepackage{appendix}
\begin{document}

\title{\boldmath Hierarchy independent sensitivity to leptonic $\delta_{CP}$ with atmospheric neutrinos}

\author{D.~Indumathi}
\email{indu@imsc.res.in}
\affiliation{The Institute of Mathematical Sciences, Chennai 600 113, India}
\affiliation{Homi Bhabha National Institute, Training School Complex, Anushakti Nagar,
     Mumbai 400085, India}

\author{S.~M.~Lakshmi}
\email{slakshmi@physics.iitm.ac.in}
\affiliation{Indian Institute of Technology Madras, Chennai 600 036, India}
          
\author{M.~V.~N.~Murthy}
\email{murthy@imsc.res.in}
\affiliation{The Institute of Mathematical Sciences, Chennai 600 113, India}

\date{\today} 
\bigskip

\begin{abstract}
The Dirac leptonic CP phase $\delta_{CP}$ is one of the crucial unknown
parameters in neutrino oscillation physics. In this paper we explore the
possibility of using low energy atmospheric neutrino events to probe
$\delta_{CP}$. We show that at sub GeV energies, when the events are
binned as a function of the energy and direction of the final state
leptons, a consistent distinction between various true $\delta_{CP}$
values is obtained. We also show that at these energies there is no
sensitivity to the mass ordering/hierarchy, so that $\delta_{CP}$
can be measured without hierarchy ambiguity. In addition
a preliminary $\chi^2$ analysis of the sensitivity to $\delta_{CP}$ using
atmospheric neutrinos assuming a generic detector with perfect
separation between charged current $\nu_\mu,\overline{\nu}_\mu,\nu_e$
and $\overline{\nu}_e$ events is given.
\end{abstract}

\maketitle
\section{Introduction}\label{intro}
   The Dirac leptonic phase $\delta_{CP}$ is one of the important
unknown parameters in neutrino oscillation physics today.
   Although there is a hint that its value is approximately $\delta_{CP} \approx -145^\circ~(-76^\circ)$ for normal (inverted) 
   hierarchy \cite{esteban}, several experiments, mainly accelerator long base line (LBL) experiments are
   running/are being designed to measure this parameter precisely \cite{t2k, nova, dune}. These 
   accelerator LBL experiments will have very high sensitivity to $\delta_{CP}$ (especially DUNE) on their 
   own; however,  it is important to study the sensitivity from atmospheric neutrinos also. Several atmospheric 
   neutrino experiments are currently running \cite{SK, SK1, SK2} and are proposed \cite{ical,HK, HK-tdr,orca,orca1,pingu} to probe different neutrino 
   oscillation parameters. Although these experiments have smaller fluxes compared to the accelerator
   LBL ones, they probe a wide range of neutrino baselines and energies, $L_\nu$ and $E_\nu$, and hence can be used to 
   probe a wide variety of physics scenarios. Also atmospheric neutrino fluxes peak at sub GeV energies \cite{honda-flux, honda1, honda2}; and 
   atmospheric neutrino experiments do not require separate runs for $\nu$ and $\overline{\nu}$ unlike the accelerator LBL 
   experiments. 

   In this paper, we explore the possibility of using atmospheric neutrino events, especially those in the sub GeV 
   region ($E_\nu<1$ GeV) to probe the CP phase $\delta_{CP}$. It has been shown earlier \cite{TA-pre,TA-hie,3d-MMD,hi-mu} 
   that atmospheric neutrinos at energies above a GeV or so are completely {\em insensitive} to the CP phase and hence can 
   be used to measure the neutrino mass ordering/hierarchy (MH) {\em independent} of the CP phase. This is in contrast 
   to beam experiments where degeneracies give rise to ambiguities in the extraction of these oscillation parameters 
   so that the MH determination is entangled with and depends on the true value of $\delta_{CP}$.

In this paper, we show for the first time that both the $\nu_e$ and
$\nu_\mu$ atmospheric neutrino events at low, i.e., sub-GeV, energies
are sensitive to the CP phase. In addition, the dependence on the CP
phase (of the angular distribution) of $\nu_e$ and $\nu_\mu$ events is
such that it {\em systematically} shifts the observed event rates in
{\em opposite} directions in the two cases. Finally, this dependence on
CP is {\em independent} of the mass hierarchy. Thus we propose that a
measurement of sub-GeV atmospheric neutrino events, where event rates
are large, will be able to determine $\delta_{CP}$ cleanly, independent
of the mass hierarchy (MH). That is, atmospheric neutrino events are uniquely
positioned so that \emph{low energy sub-GeV events are sensitive to
$\delta_{CP}$ independent of MH}, while \emph{higher energy
(few GeV) events are sensitive to MH independent of $\delta_{CP}$}.
We believe that this unique dependence has been discussed for the first
time, in this paper.

In particular, we show in Section~\ref{posc-delcp} that the events
spectra for different values of the true $\delta_{CP}$ are different
when the events are binned as a function of the energy and direction
of the final state lepton in the charged current (CC) interaction,
thus making it possible to have a good $\delta_{CP}$ sensitivity. In
Section.~\ref{analytic} we show analytically, that at such low energies
and larger baselines which are relevant for atmospheric neutrinos,
$\delta_{CP}$ can be determined {\em irrespective of hierarchy}. We
quantify our results in Section.~\ref{chi2} with a preliminary and simple
$\chi^2$ analysis of the events that would be obtained with a detector
with perfect resolutions and efficiencies, and ignoring systematic
effects. We end with discussions and conclusions which are presented
in Section.~\ref{conclusions}.

 \section{Oscillation probabilities and atmospheric neutrino events at sub GeV energies}\label{posc-delcp}

\begin{figure}[htp] \centering
\includegraphics[width=0.45\textwidth,height=0.45\textwidth]{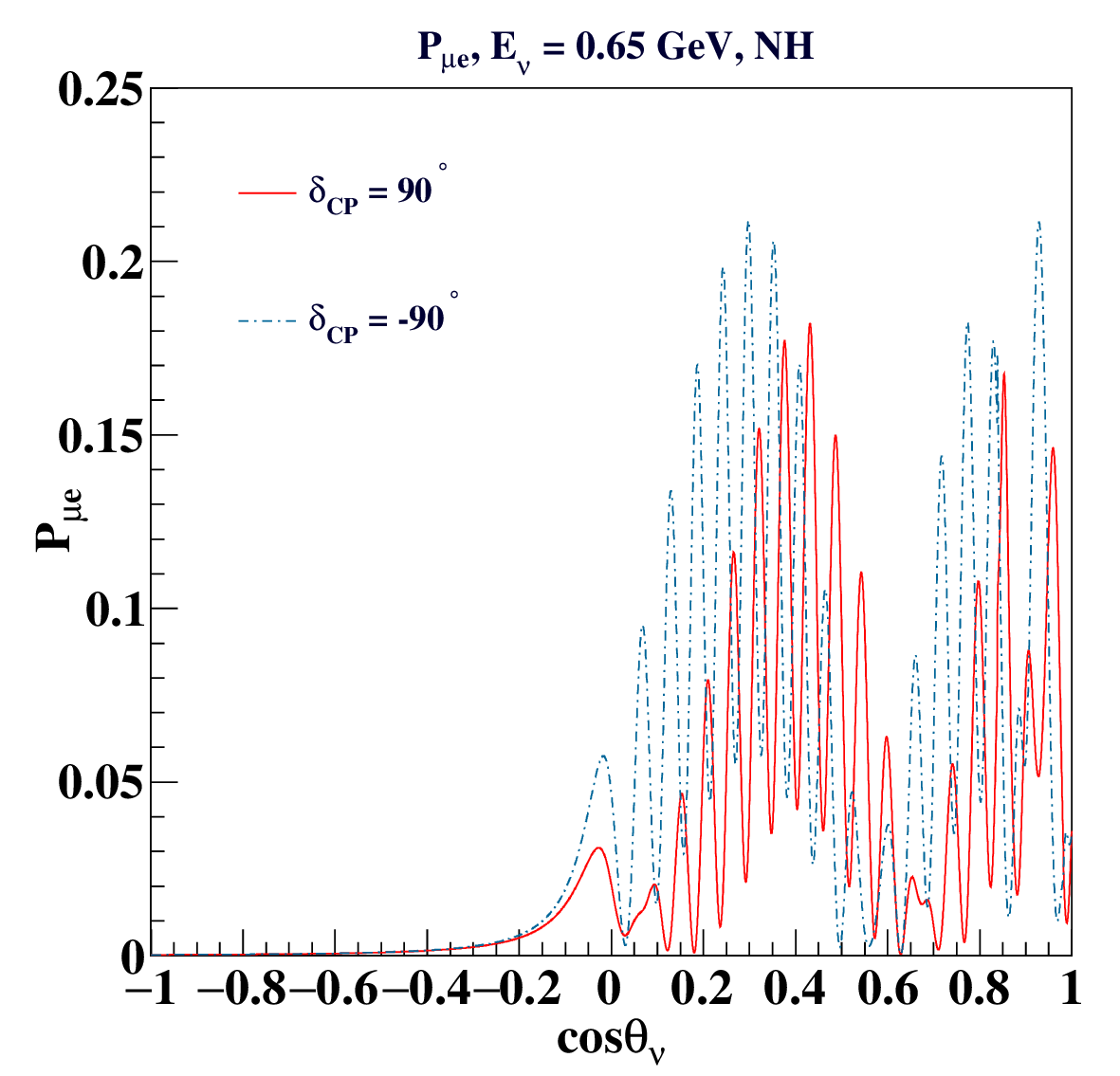}
\includegraphics[width=0.45\textwidth,height=0.46\textwidth]{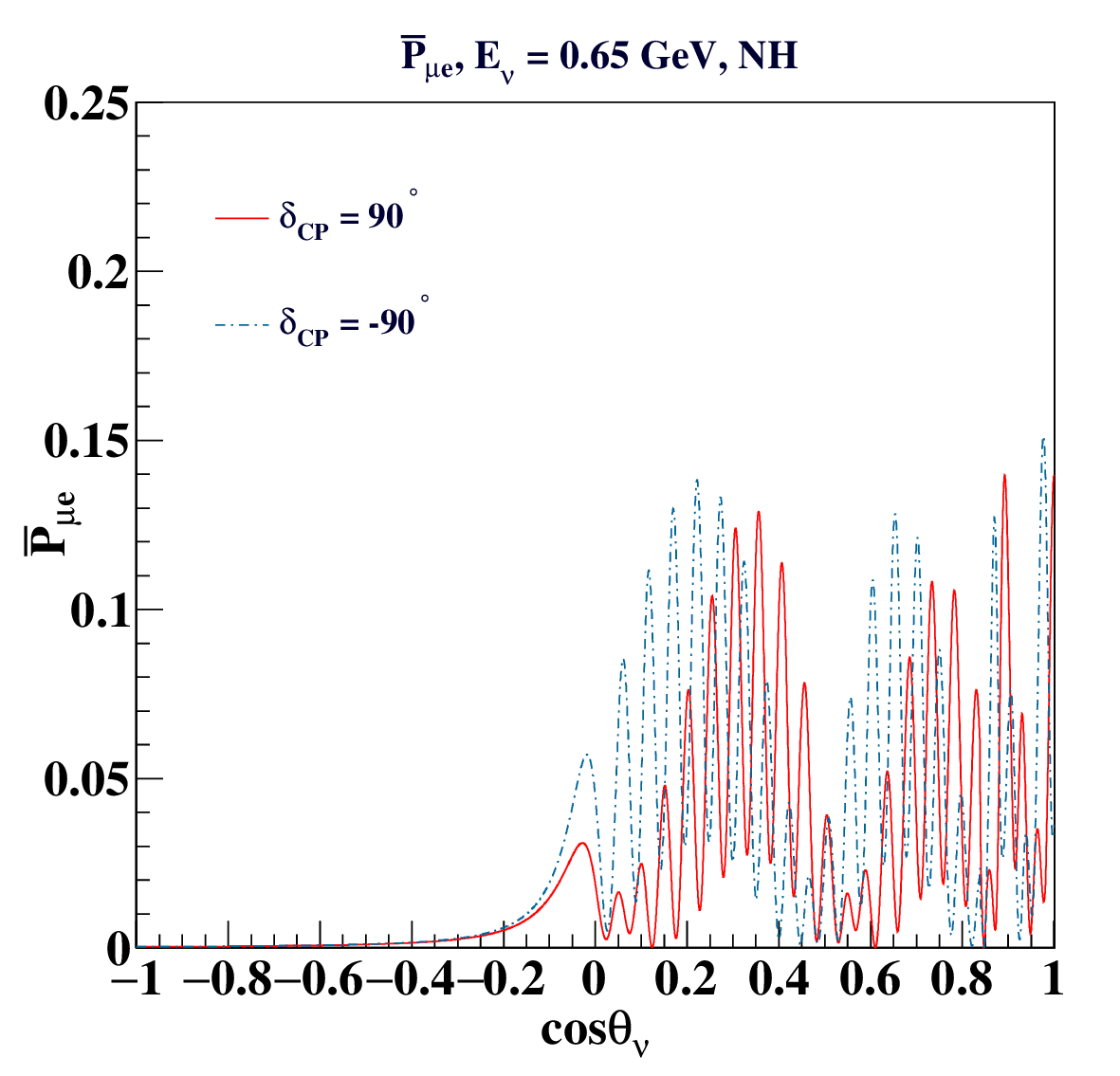}

\includegraphics[width=0.45\textwidth,height=0.45\textwidth]{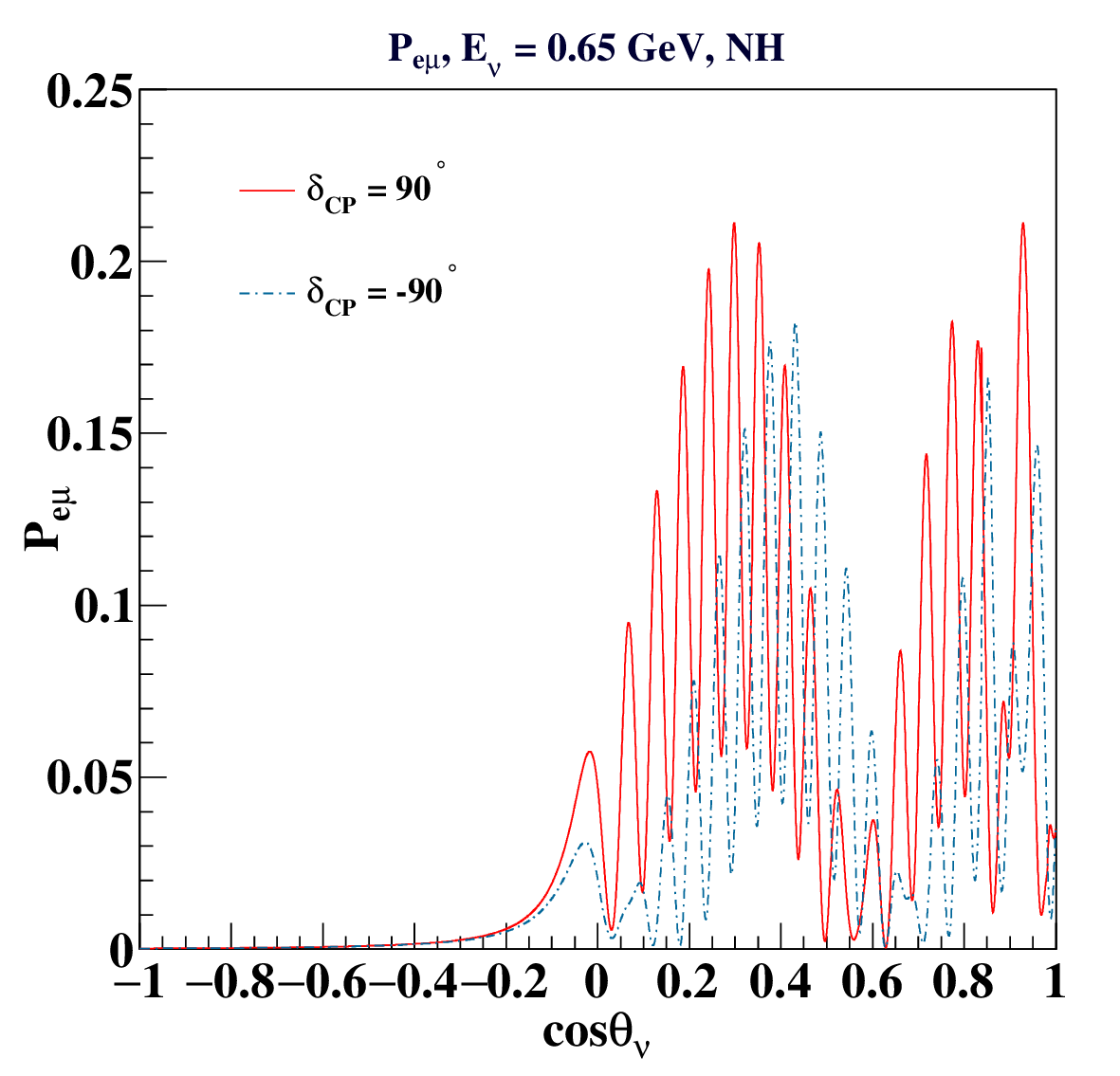}
\includegraphics[width=0.45\textwidth,height=0.45\textwidth]{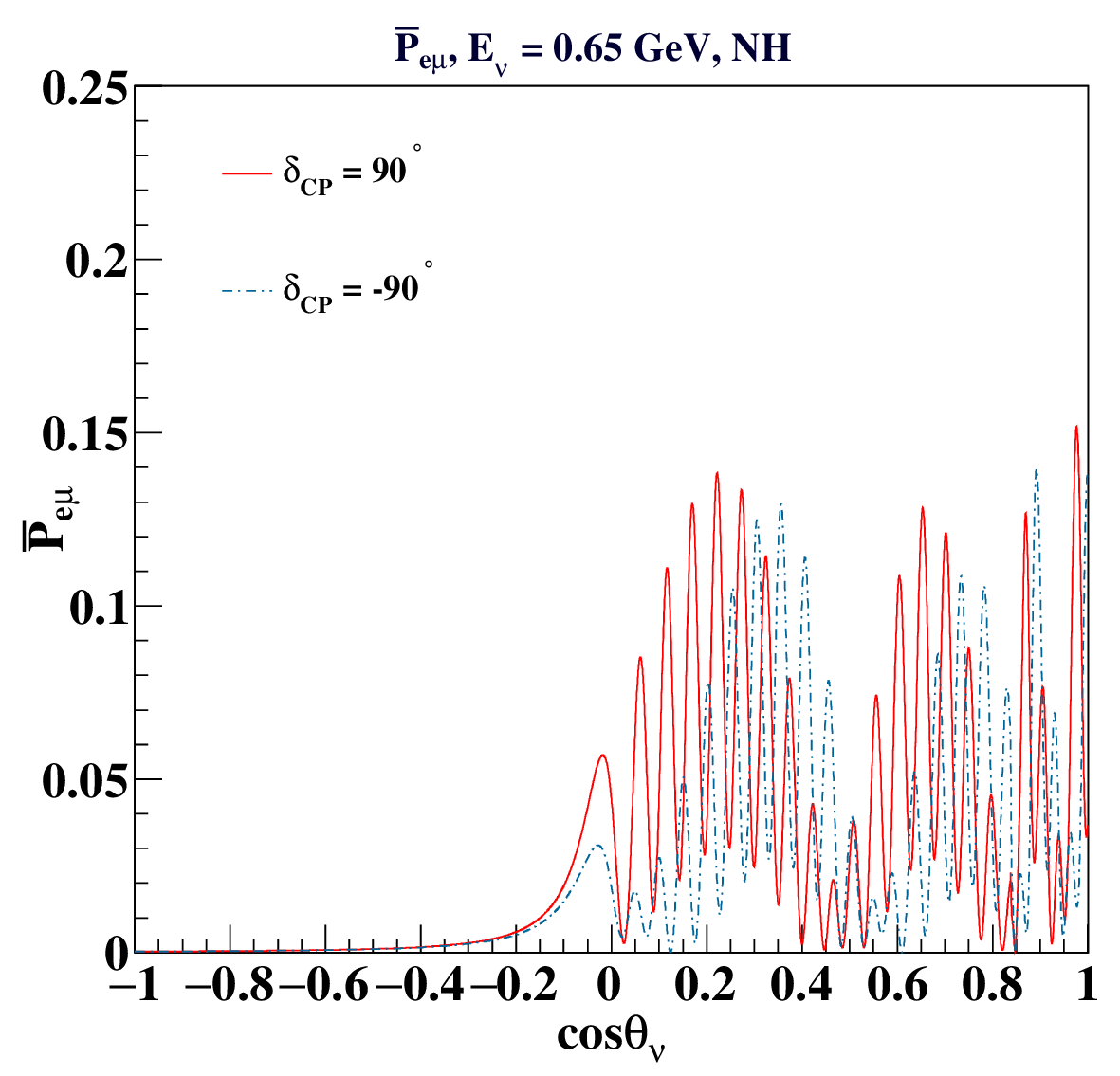}
\caption{Transition probabilities $P_{\mu e}$ (top-left),
$\overline{P}_{\mu e}$ (top-right), $P_{e\mu}$ (bottom-left) and
$\overline{P}_{e\mu}$ (bottom-right) as a function of $\cos\theta_\nu$ for a
fixed neutrino energy, $E_\nu$ = 0.65 GeV and true normal mass hierarchy.}
\label{posc-0.65-GeV}
\end{figure}

In the case of atmospheric neutrinos, the muon neutrino (and
anti-neutrino) fluxes  $\Phi_\mu+\overline{\Phi}_\mu$ are larger that of the
electron neutrino (and anti-neutrino) fluxes $\Phi_e+\overline{\Phi}_e$.
The sensitivity to $\delta_{CP}$ in these events arises from
the transition probabilities $P_{\mu e},\overline{P}_{\mu e}$
and $P_{e\mu},\overline{P}_{e\mu}$ \cite{kimura1, minako-honda, krastev}. Hence the
sensitivity to $\delta_{CP}$ will be larger for electron like events as
compared to the muon like events. A comparison of the relevant oscillation probabilities for two
different values of $\delta_{CP}$ as a function of $\cos\theta_\nu$
(neutrino direction) for a sample sub-GeV neutrino energy, $E_\nu$ =
0.65 GeV is shown in Fig.~\ref{posc-0.65-GeV}.

The probabilities are flipped for $P_{\mu{e}}$ ($\overline{P}_{\mu{e}}$)
and $P_{e\mu}$ ($\overline{P}_{\mu{e}}$) when $\delta_{CP}$ changes from
$90^\circ\leftrightarrow-90^\circ$. In each case, the probabilities with
one value of $\delta_{CP}$, say $90^\circ$ is not always below that with
$-90^\circ$,  i.e., the trend is not consistently larger or smaller in
all bins, but shows an oscillatory pattern.

\begin{figure}[htp] \centering
\includegraphics[width=0.45\textwidth,height=0.45\textwidth]{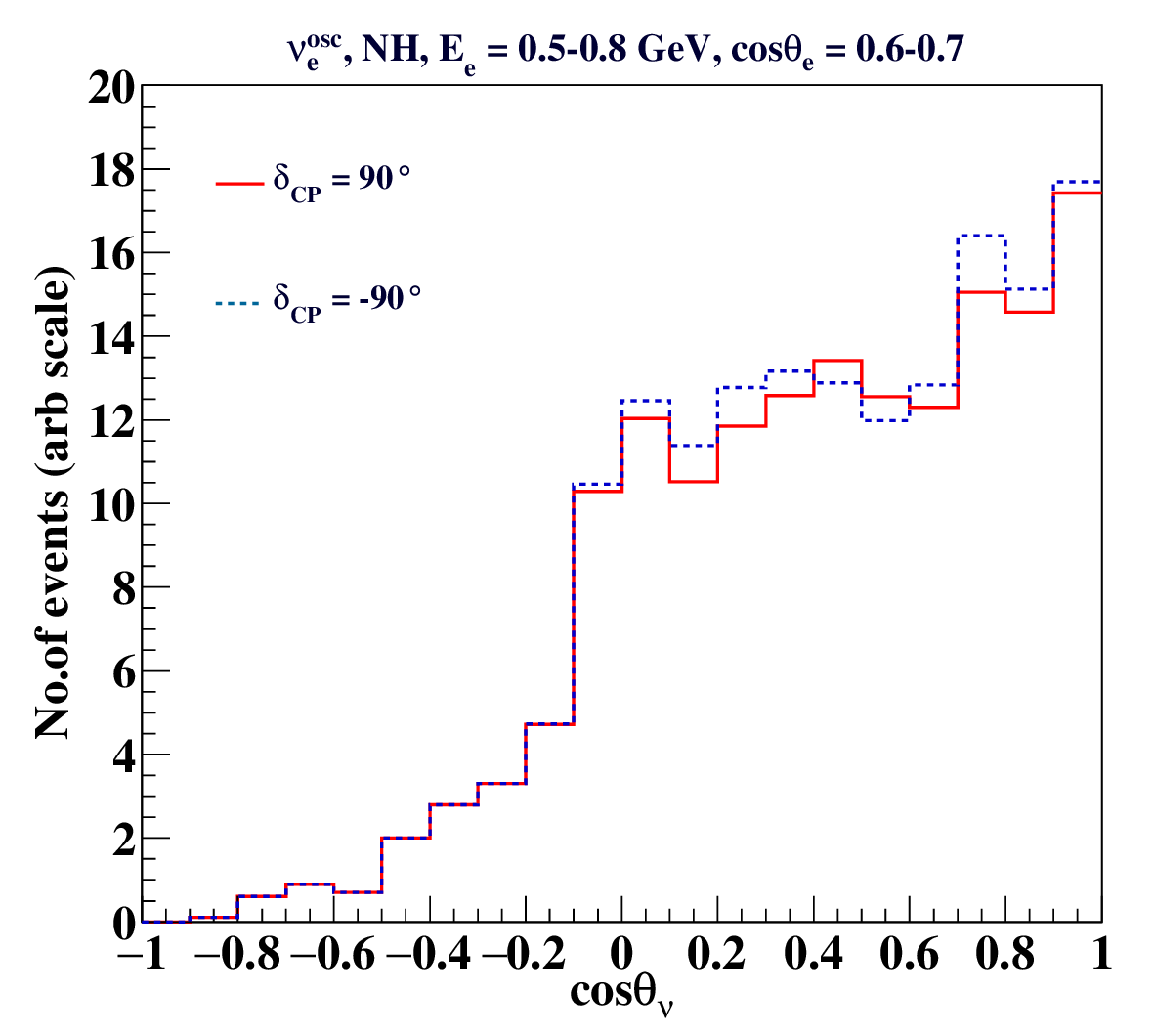}
\includegraphics[width=0.45\textwidth,height=0.45\textwidth]{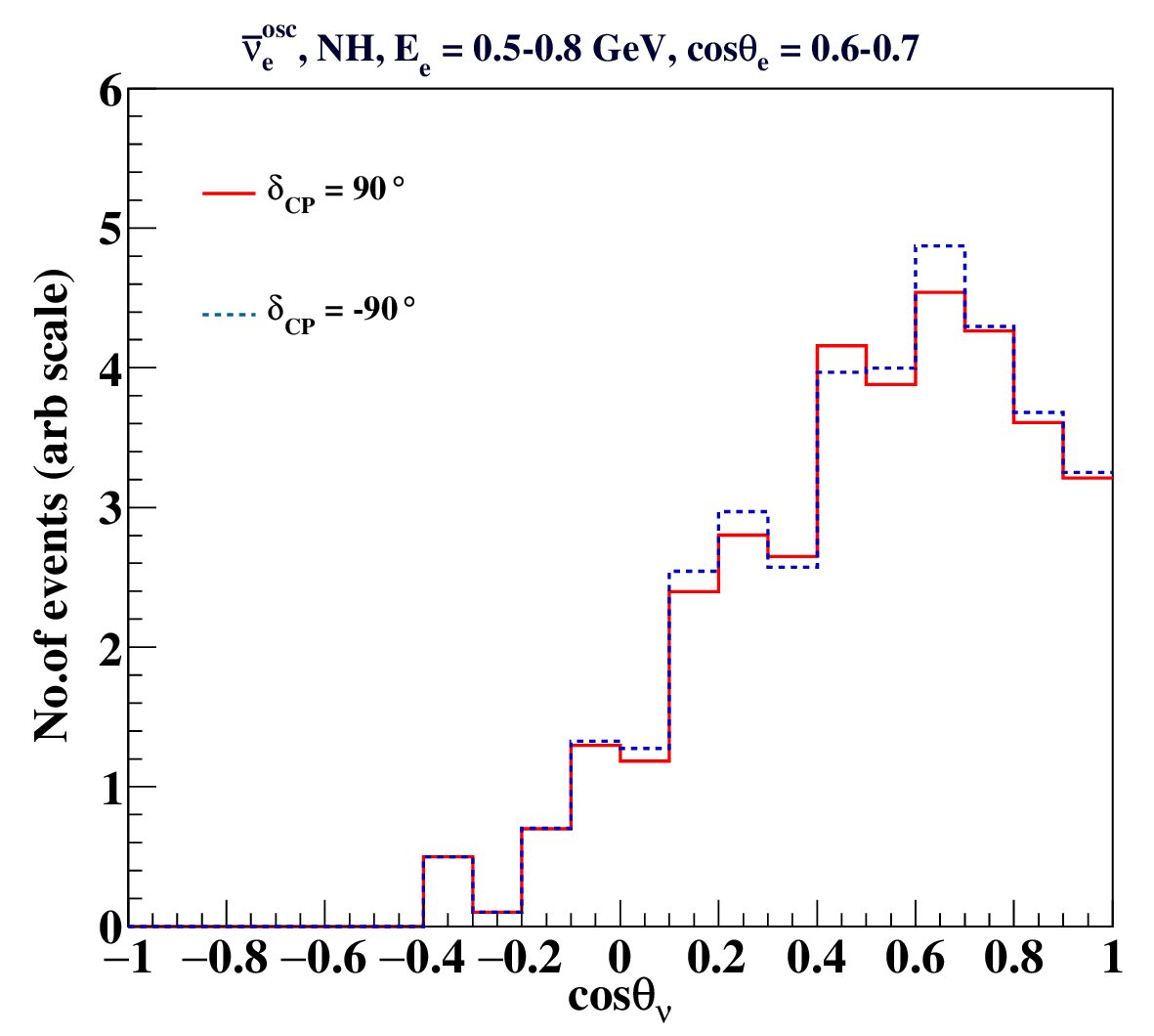}

\includegraphics[width=0.45\textwidth,height=0.445\textwidth]{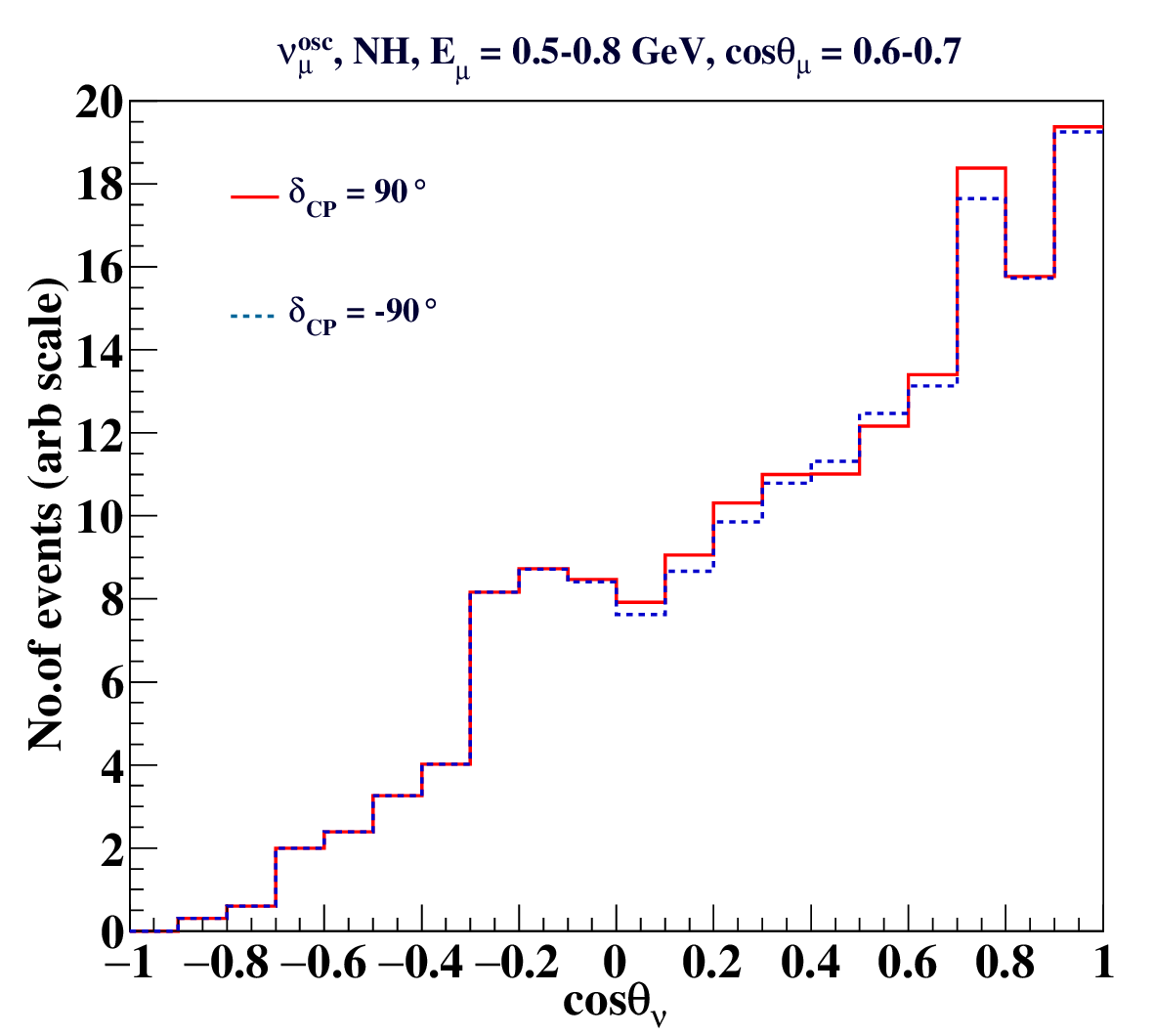}
\includegraphics[width=0.45\textwidth,height=0.445\textwidth]{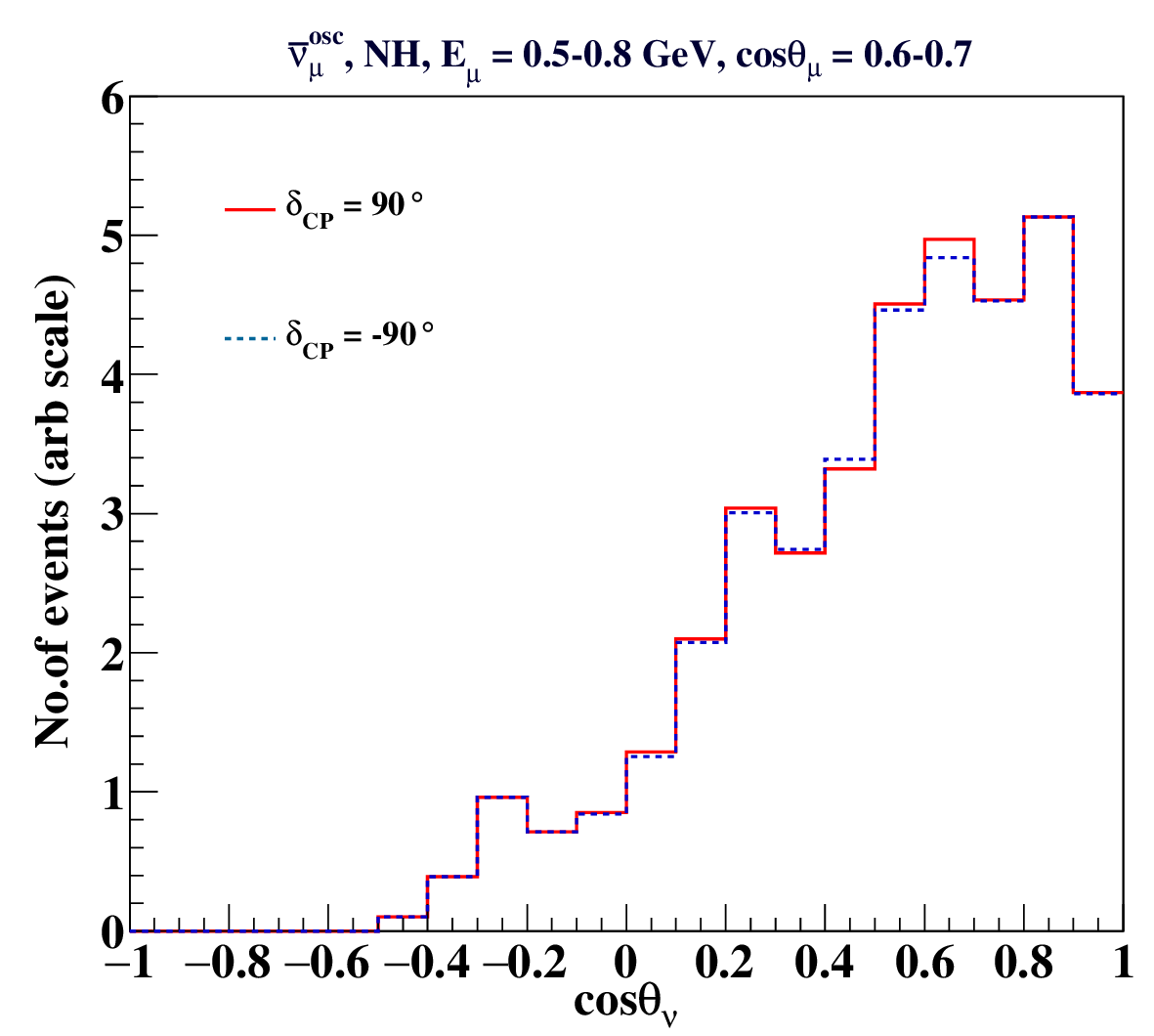}
\caption{ (Top) Oscillated electron type; (bottom) muon type events
events as a function of neutrino direction, $\cos\theta_\nu$,
for events with final lepton energy, $E_l$ = 0.5--0.8 GeV and direction 
$\cos\theta_l$ = 0.6--0.7, with $\delta_{CP}=\pm90^\circ$ and true NH. 
The left panels are for $\nu$ events and the right ones are for 
$\overline{\nu}$ events.}
\label{eve-ctnu}
\end{figure}

\begin{figure}[htp] \centering
\includegraphics[width=0.45\textwidth,height=0.45\textwidth]{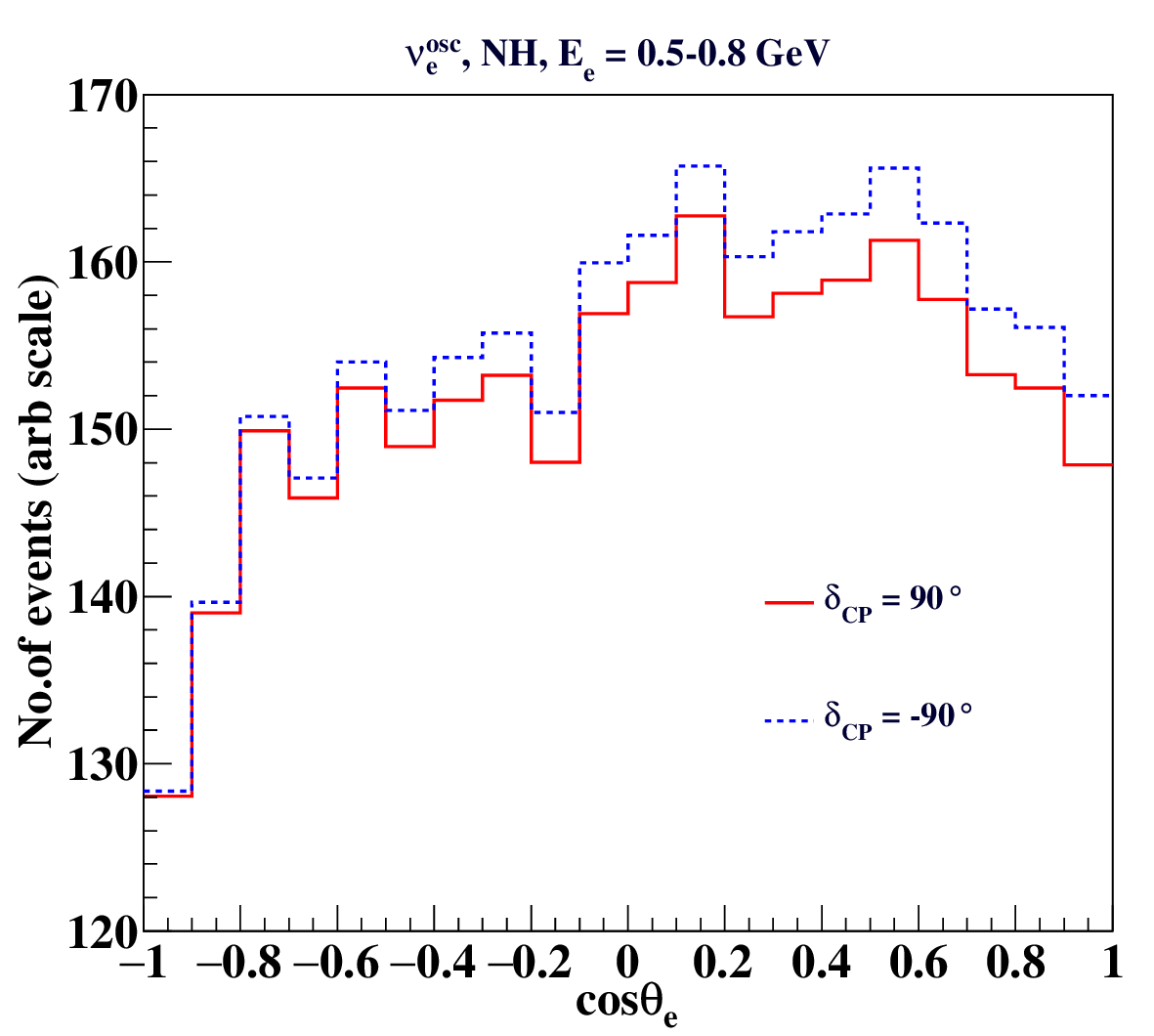}
\includegraphics[width=0.45\textwidth,height=0.45\textwidth]{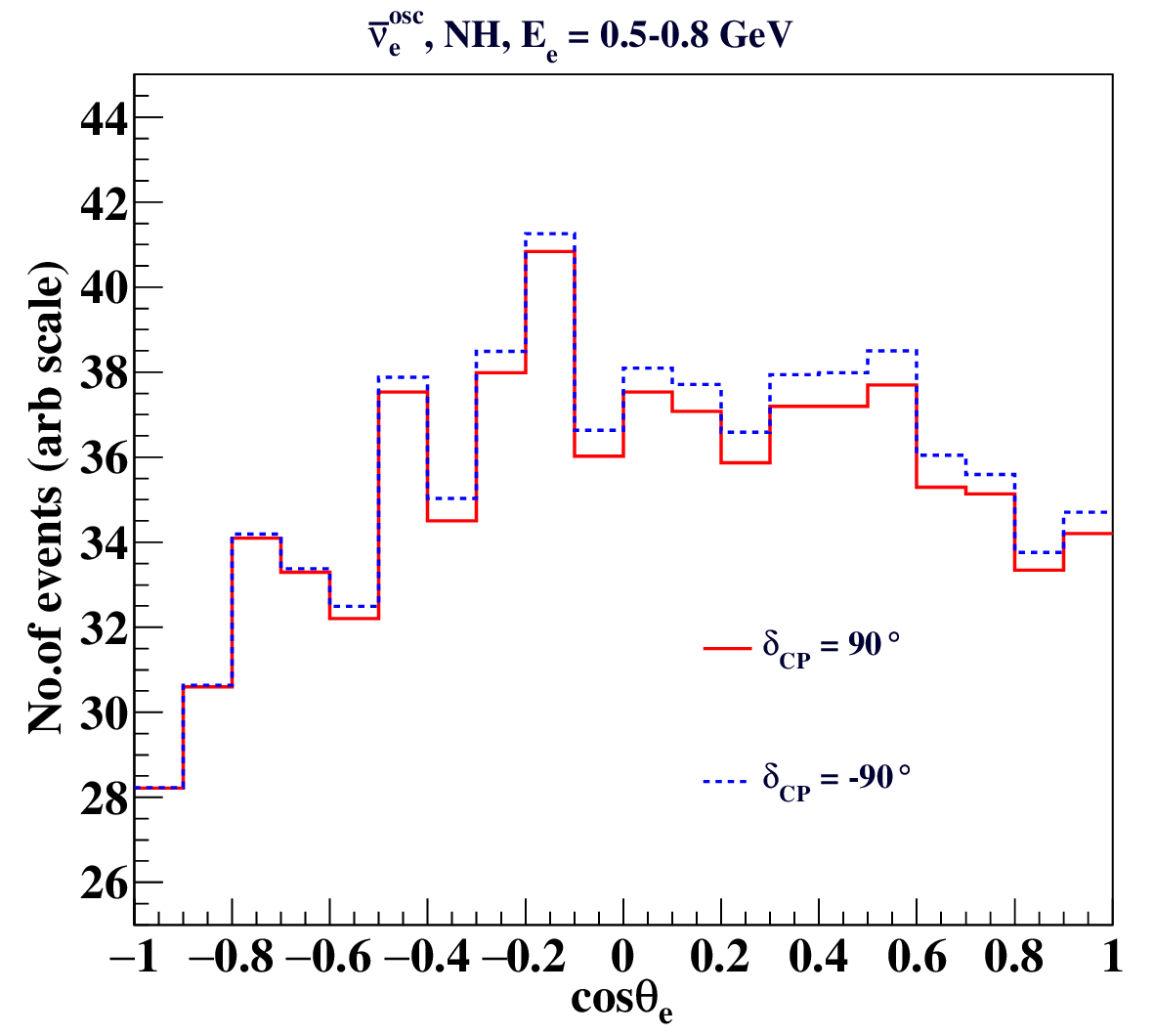}

\includegraphics[width=0.45\textwidth,height=0.445\textwidth]{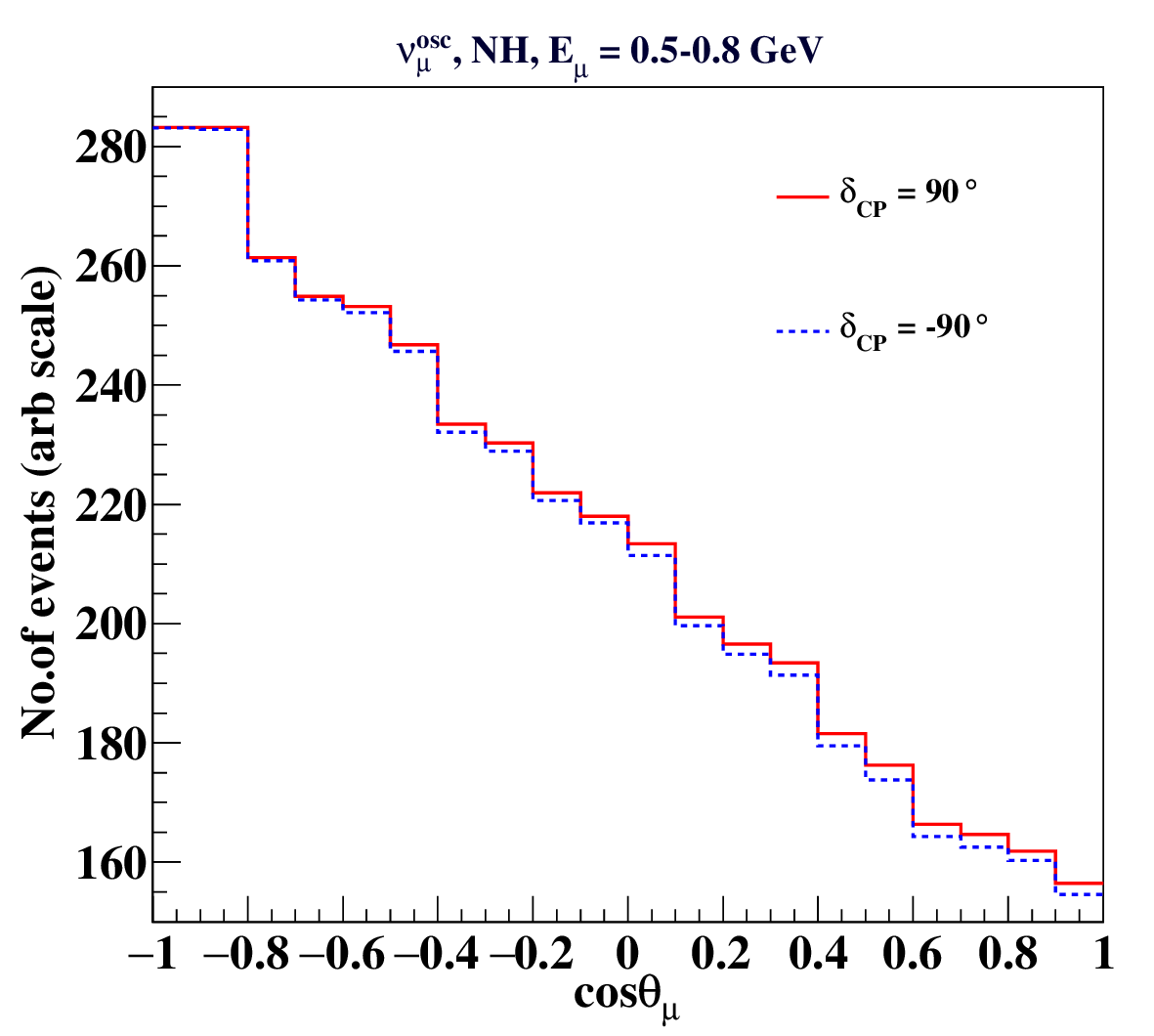}
\includegraphics[width=0.45\textwidth,height=0.445\textwidth]{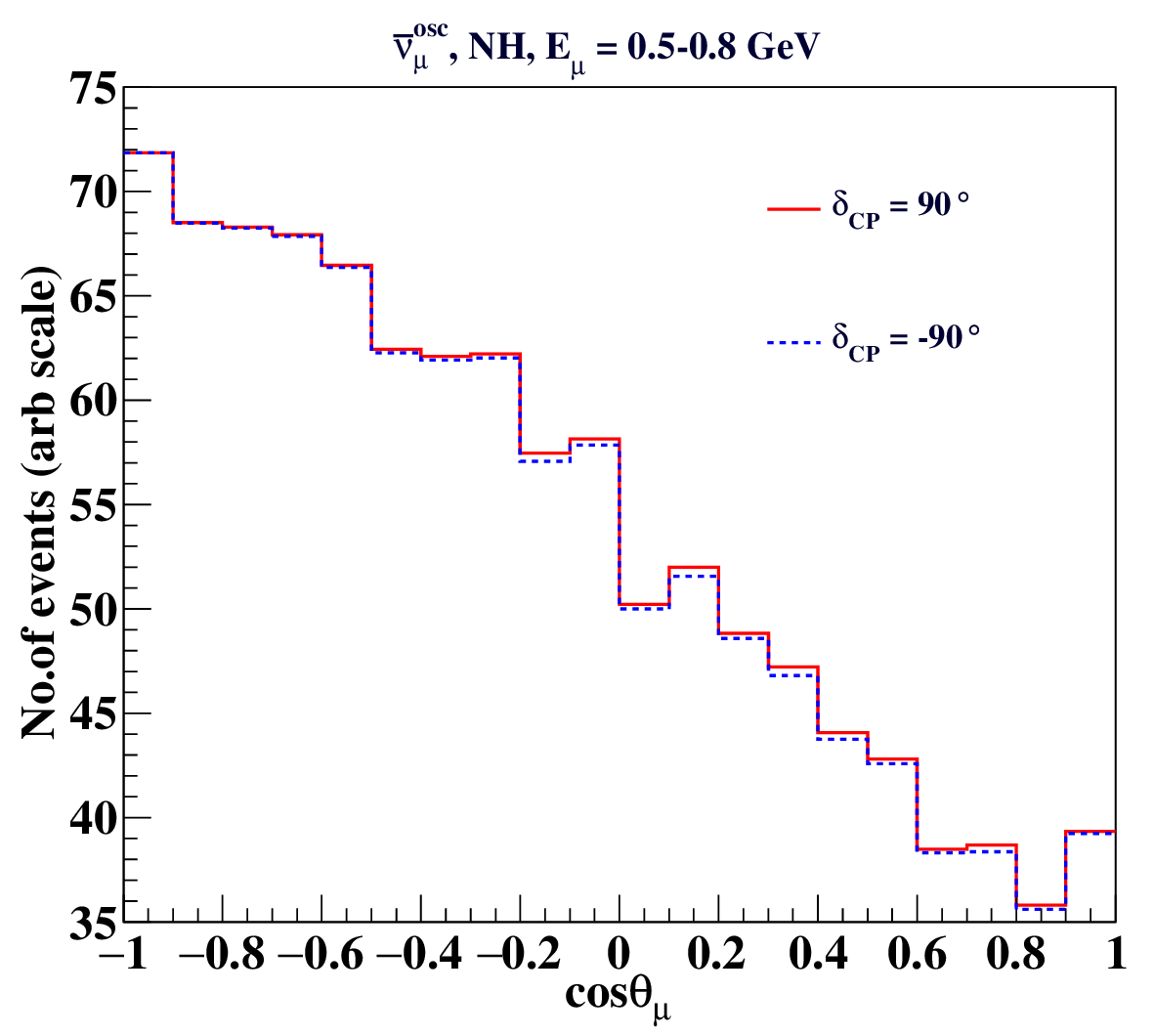}
\caption{ (Top) Oscillated electron type; (bottom) muon type events
events as a function of the final state lepton angle, $\cos\theta_l$,
for events with final lepton energy, $E_l$ = 0.5--0.8 GeV, with
$\delta_{CP}=\pm90^\circ$ and true NH. The left panels are for $\nu$
events and the right ones are for $\overline{\nu}$ events. Note that all y-axes scales are different.}
\label{evts-ctl-0.65GeV}
\end{figure}

A similar behaviour can be seen if we plot the events spectra as a
function of $\cos\theta_\nu$, as in Fig.~\ref{eve-ctnu}. The events
are generated with parameters given in Table.~\ref{osc-par-3sig}. Here
the oscillated events are plotted as a function of $\cos\theta_\nu$
for bins of final state lepton energy and angle, $E_l = 0.5$--0.8 GeV
and $\cos\theta_l = 0.6$--0.7, where $l=e,\mu$. The event spectrum follows
the oscillatory behaviour of the transition probabilities plotted in
Fig.~\ref{posc-0.65-GeV}, where some bins have more events when
$\delta_{CP}=+90^\circ$ than $-90^\circ$, while it is reversed in other
bins. But when we plot the events as a function of the lepton direction
$\cos\theta_l$, which is the true observable, the behaviour changes as
shown in Fig.~\ref{evts-ctl-0.65GeV}.

It can be seen from Fig.~\ref{evts-ctl-0.65GeV} that when the oscillated
events are plotted as a function of $\cos\theta_l$, the spectrum with
$\delta_{CP}=-90^\circ$ is always greater (less) than that with $90^\circ$
for $\nu_e$ and $\overline{\nu}_e$ ($\nu_\mu$ and $\overline{\nu}_\mu$), although
the effect is smaller for muon neutrinos. That is, the ``oscillatory''
dependence on $\delta_{CP}$ seen in Figs.~\ref{posc-0.65-GeV} and
\ref{eve-ctnu} has disappeared, giving rise to a {\em systematic
dependence} on the CP phase. This effect is due to the kinematics of the
interaction which generates a final state lepton scattered at an angle
that can be far different from that of the parent neutrino. This is
especially so at low energies of interest here, where the dominant
process is quasi-elastic (QE) neutrino-nucleus scattering.

Thus, when plotted as a function of $\cos\theta_l$, the features of
the underlying oscillation probability are lost but the sensitivity
to $\delta_{CP}$ is reinforced. This behaviour, arising because of the
different distributions of the final state lepton due to the kinematics of
the interaction has not been discussed in the literature before. Again,
the same systematic behaviour is also seen in muon events. Although
the effect is weaker, $\delta_{CP}$ dependence is opposite to that of
electron events.

\begin{figure}[htp] \centering
\includegraphics[width=0.45\textwidth,height=0.45\textwidth]{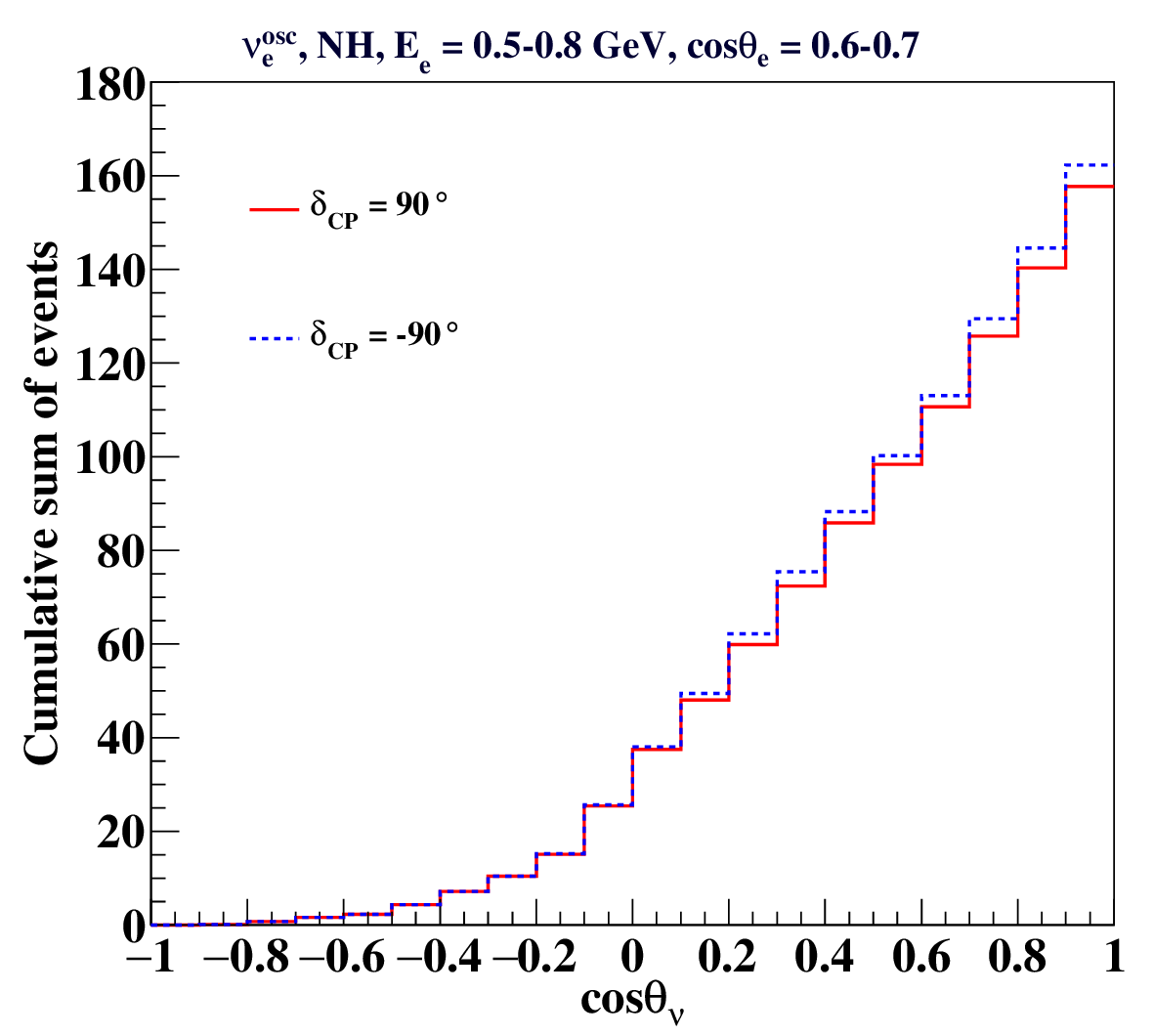}
\includegraphics[width=0.45\textwidth,height=0.45\textwidth]{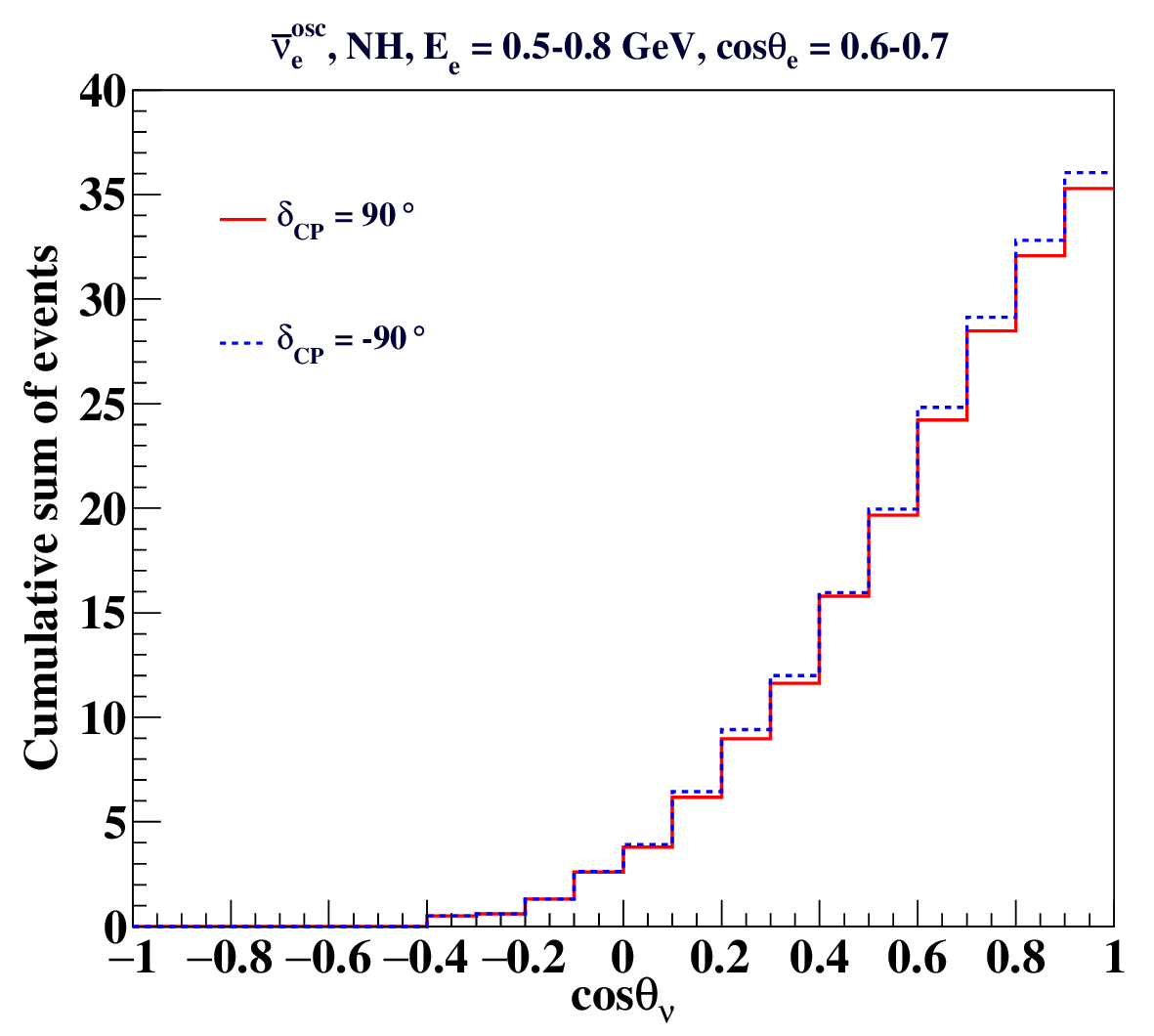}

\includegraphics[width=0.45\textwidth,height=0.445\textwidth]{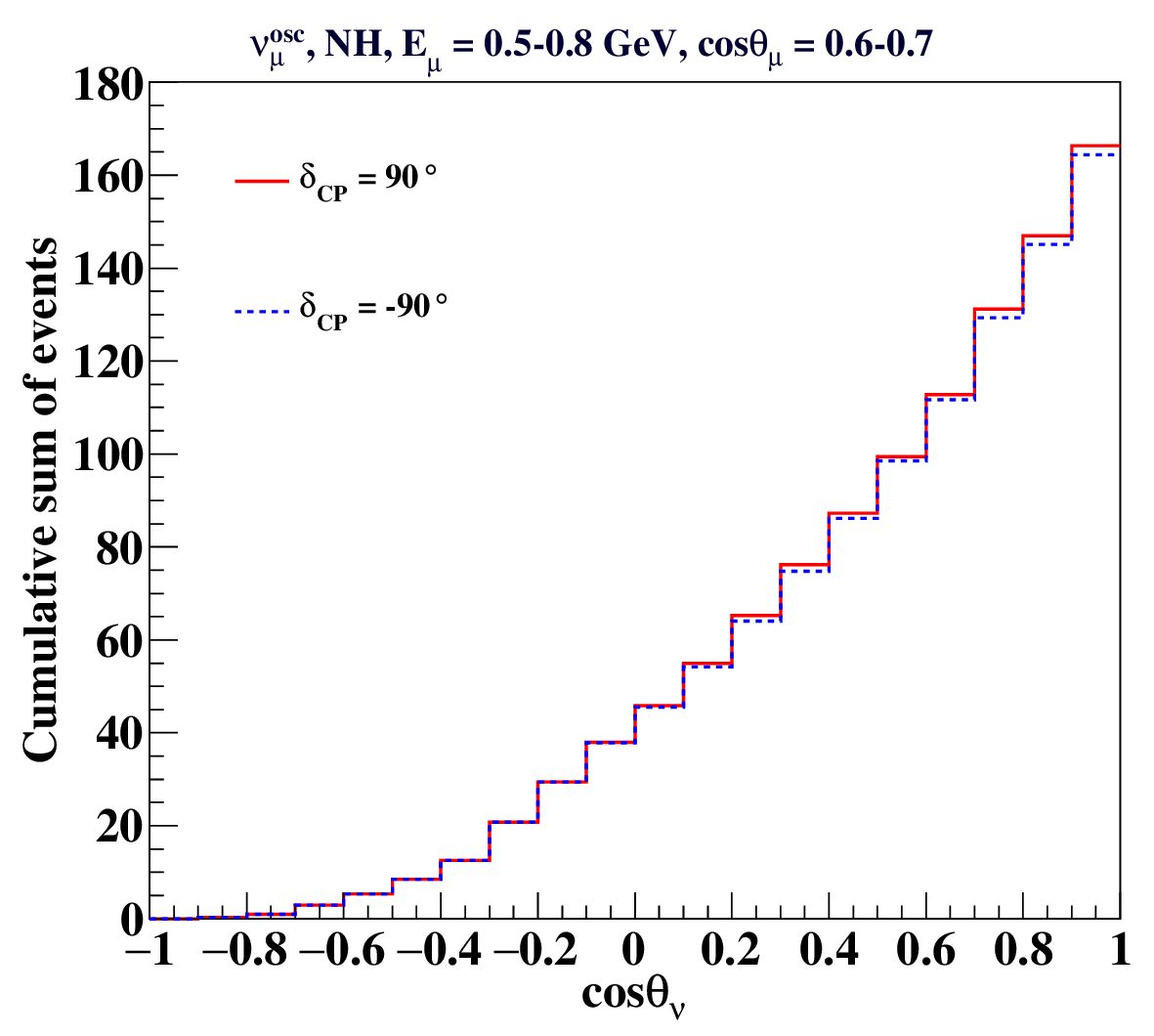}
\includegraphics[width=0.45\textwidth,height=0.445\textwidth]{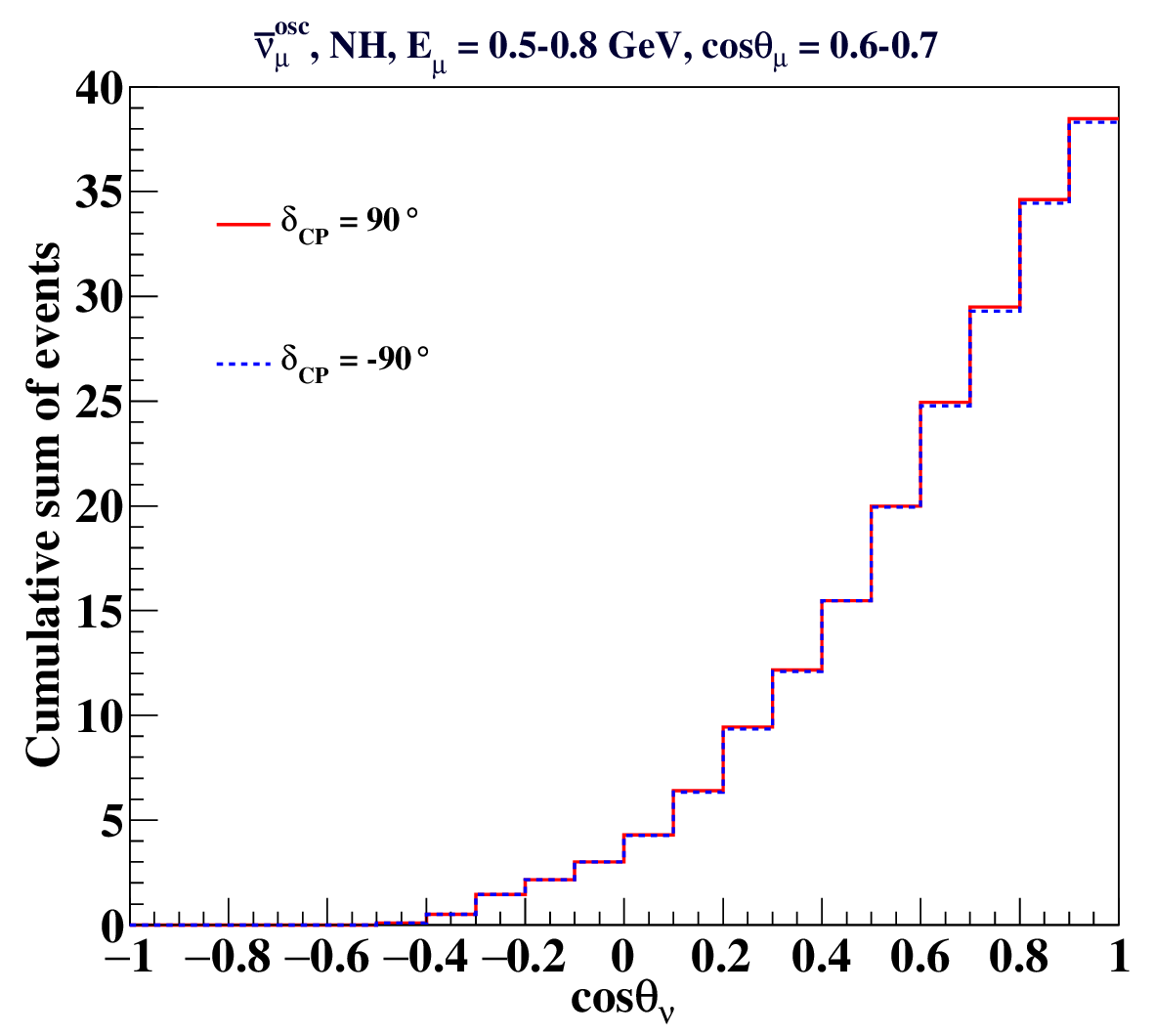}
\caption{Cumulative sum of oscillated (top) electron type (bottom)
muon type events, which contribute to the $\cos\theta_l$ bin 0.6--0.7,
as a function of $\cos\theta_\nu$ for the bin $E_l$ = 0.5--0.8 GeV with
$\delta_{CP}=\pm90^\circ$ and true NH. The last bin in each plots gives
the total contribution to each $\cos\theta_l$ bin. The left panels
are for $\nu$ events and the right ones are for $\overline{\nu}$ events.}
\label{cum-sum-ctnu}
\end{figure}

The cumulative sum of events as a function of $\cos\theta_\nu$ is shown
in Fig.~\ref{cum-sum-ctnu} to illustrate this. It can be seen that
though the actual distribution of events in $\cos\theta_\nu$ follows the
oscillation probabilities, scattering of neutrinos coming from several
neutrino directions $\cos\theta_\nu$, give rise to leptons with the same
scattering angle, $\cos\theta_l$. For instance, neutrinos with
$\cos\theta_\nu$ practically from $-1$ to $+1$ contribute to the events
in the bin with $0.6 \le \cos\theta_l \le 0.7$.

When we sum over these events, their contributions to each $\cos\theta_l$
bin do not resemble the probabilities, but follow a pattern in which
the events spectrum with a particular value of $\delta_{CP}$ is always
greater than the other because of the kinematics of the events. In
addition, we have shown all plots so far assuming the normal hierarchy.
We will now show, both analytically and numerically, that these results
hold, irrespective of the MH.

\section{Hierarchy (in)dependence at low energies}
\label{analytic}

In this section we show that at low energies there is no hierarchy
ambiguity for atmospheric neutrinos and hence $\delta_{CP}$ can
be measured irrespective of hierarchy. In fact, this can be established analytically
as we show here.

\subsection{Hierarchy independence: analytic approach}

The 3-flavour vacuum oscillation probability of a flavour
$\nu_\alpha\to\nu_\beta$ is given by :
\begin{eqnarray}
\overset{(-)}P^{vac}_{\alpha\beta}&=&\delta_{\alpha\beta}-4\sum_{i>j}Re\left[U_{\alpha{i}}U^*_{\beta{i}}U^*_{\alpha{j}}U_{\beta{j}}\right]
\sin^2\left(\frac{1.27\Delta{m^2_{ij}L}}{E}\right)\\
\nonumber {\hspace{-30ex}} &\pm&
2\sum_{i>j}Im\left[U_{\alpha{i}}U^*_{\beta{i}}U^*_{\alpha{j}}U_{\beta{j}}\right]\sin\left(\frac{2.53\Delta{m^2_{ij}}L}{E}\right),
\label{pab}
\end{eqnarray}
where $\alpha,\beta=e,\mu,\tau$ are the flavour
indices, and the $\pm$ sign corresponds to neutrinos and anti-neutrinos
respectively.

Here
$$
U^{vac}_{\alpha{i}} = \begin{pmatrix}
c_{12}c_{13} & s_{12}c_{13} & s_{13}e^{-i\delta} \\
-c_{23}s_{12}-s_{23}c_{12}s_{13}e^{i\delta} & c_{23}c_{12}-s_{23}s_{12}s_{13}e^{i\delta} & s_{23}c_{13} \\
s_{23}s_{12}-c_{23}c_{12}s_{13}e^{i\delta} & -s_{23}c_{12}-c_{23}s_{12}s_{13}e^{i\delta} & c_{23}c_{13}
\end{pmatrix}~,
$$
where $c_{ij}~=~\cos\theta_{ij}$, $s_{ij}~=~\sin\theta_{ij}$; $i,j=1,2,3$
are the mass eigenstates, $\Delta{m^2_{ij}}=m^2_i-m^2_j$ ($j<i$),
$\theta_{ij}$ are the mixing angles and $\delta_{CP}$ is the leptonic
CP violation phase. Here $L$ (in km) is the distance travelled by a
neutrino of energy $E$ (in GeV).

The survival probability $P_{\alpha\alpha}$ has no imaginary part,
while for transition probabilities $\alpha\neq\beta$, the imaginary
part changes sign with $P_{\alpha\beta}=\overline{P}_{\beta\alpha}$,
the corresponding antineutrino probability.  When $E$ is
small, of the order of a few hundred MeV, the corresponding
oscillatory terms average out whenever $L/E$ is large compared to
$\Delta{m^2_{ij}}$. Since $|\Delta{m^2_{3j}}|\sim2.4\times10^{-3}~eV^2
\gg \Delta{m^2_{21}}\sim7.6\times10^{-5}~eV^2$, $j=1,2$; this applies
to the ``atmospheric'' terms :
\begin{equation}
1.27\Delta m^2_{3j}~\frac{L}{E} \approx \pi
 \frac{(L/100~\rm{km})}{(E/0.1~\rm{GeV})}~,
\end{equation}
rather than to ``solar'' terms :
\begin{equation}
1.27\Delta m^2_{21}~\frac{L}{E} \approx \pi
 \frac{(L/3000~\rm{km})}{(E/0.1~\rm{GeV})}~.
\end{equation}
It immediately follows that the atmospheric event rates at these low
energies and for $L\geq$ a few 100 km become independent 
of $\Delta{m^2_{32}}$ and $\Delta{m^2_{31}}$ and hence of their ordering. 
The solar mass-squared difference remains, but its magnitude and sign are well known. Hence the
CP phase dependence can be studied with low energy atmospheric
neutrinos, independent of the MH. In particular, as has been discussed in the literature earlier
(see, for example, Ref.~\cite{kimura1,kimura}), the probabilities involving $e$ and $\mu$ are linear in 
$\sin\delta_{CP}$ and $\cos\delta_{CP}$. The survival probabilities are independent of $\sin\delta_{CP}$ which 
occurs in the imaginary part of the transition probabilities. In fact, $P_{ee}$ is independent of
both $\sin\delta_{CP}$ and $\cos\delta_{CP}$ while $P_{\mu\mu}$ depends on $\cos\delta_{CP}$ and
$\cos2\delta_{CP}$. The transition probabilities $P_{e\mu}$ and
$P_{\mu e}$ are thus most sensitive to $\delta_{CP}$, measurable
in principle, via a CP asymmetry, that can be expressed in
vacuum as,
\begin{eqnarray}
A_{CP}=\frac{P_{e\mu}-P_{\mu e}}{P_{e\mu}+P_{\mu e}} = -\frac{C}{A+B\cos\delta}\sin\delta \\  
\overline{A}_{CP} = \frac{\overline{P}_{e\mu}-\overline{P}_{\mu e}}{\overline{P}_{e\mu}+\overline{P}_{\mu e}} = \frac{C}{A+B\cos\delta}\sin\delta
\end{eqnarray}
for $\nu$ and $\overline{\nu}$ respectively. In matter, $A,B,C$ are modified
according to Earth matter effects on the oscillation parameters. The
linear dependence on $\cos\delta_{CP}$ and $\sin\delta_{CP}$ remains
unaltered. See Appendix \ref{appendix:A} for details.

\subsection{Hierarchy independence: Events spectra at low energies}\label{events}

It is known that Earth matter resonance occurs in atmospheric neutrinos
at a few GeV energies thus enabling the determination of the neutrino
mass hierarchy. The advantage of using atmospheric neutrinos for
hierarchy determination is that it can be determined \emph{unambiguous}
of $\delta_{CP}$, especially using the $P_{\mu\mu}$ ($\overline{P}_{\mu\mu}$)
survival channel. We have now shown that, at lower energies (sub GeV
range) this effect is reversed, i.e., $\delta_{CP}$ can be determined
irrespective of the hierarchy. In addition, we examined the CP
sensitivity of the events as a function of the final lepton scattering
angle, $\cos\theta_l$. We now show that the systematic dependence on
$\delta_{CP}$ remains when we integrate out the anglular dependence and
examine the events as a function of the final state lepton energy, $E_l$ alone.

\begin{figure}[htp] \centering
\includegraphics[width=0.45\textwidth,height=0.45\textwidth]{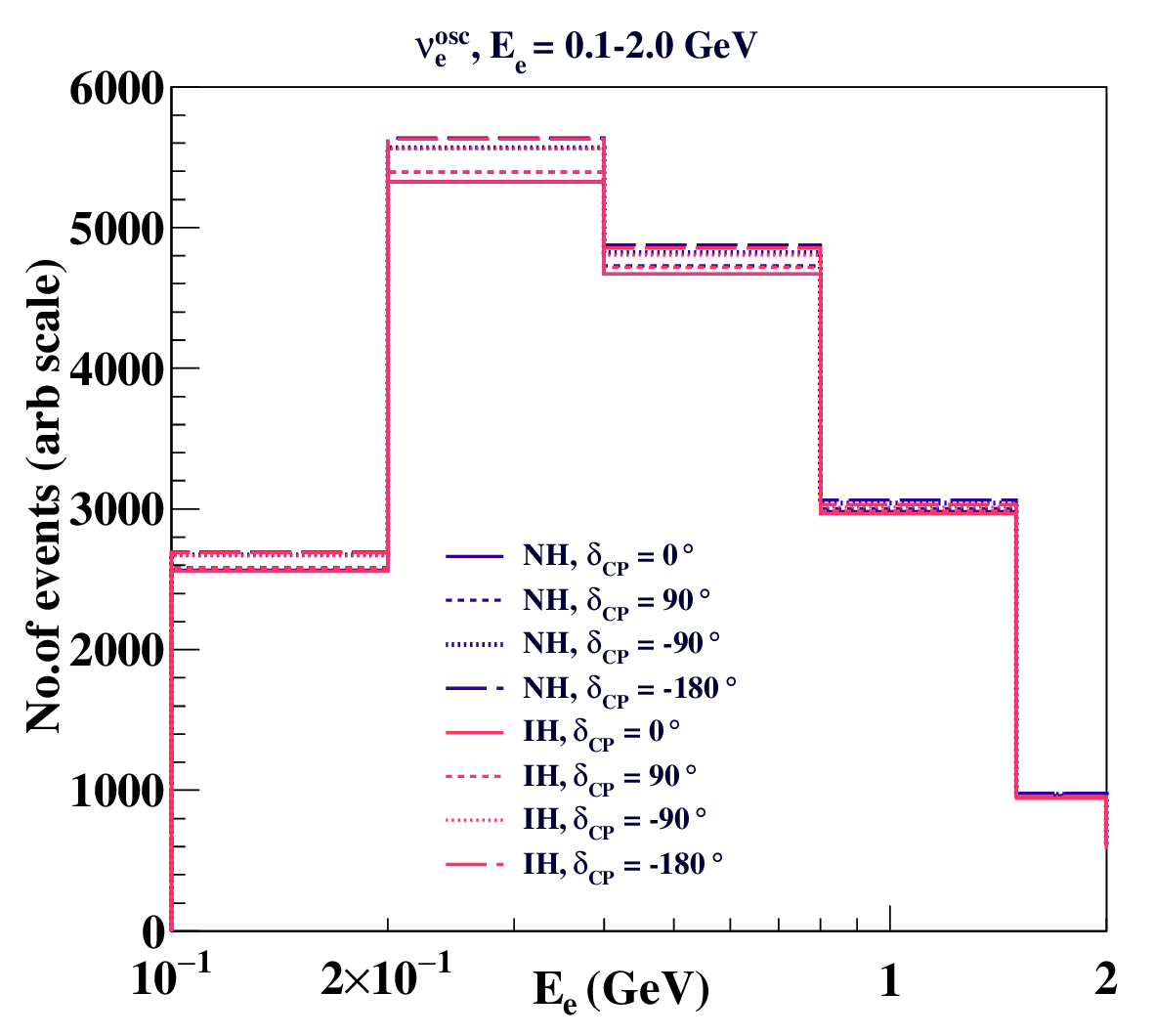}
\includegraphics[width=0.45\textwidth,height=0.45\textwidth]{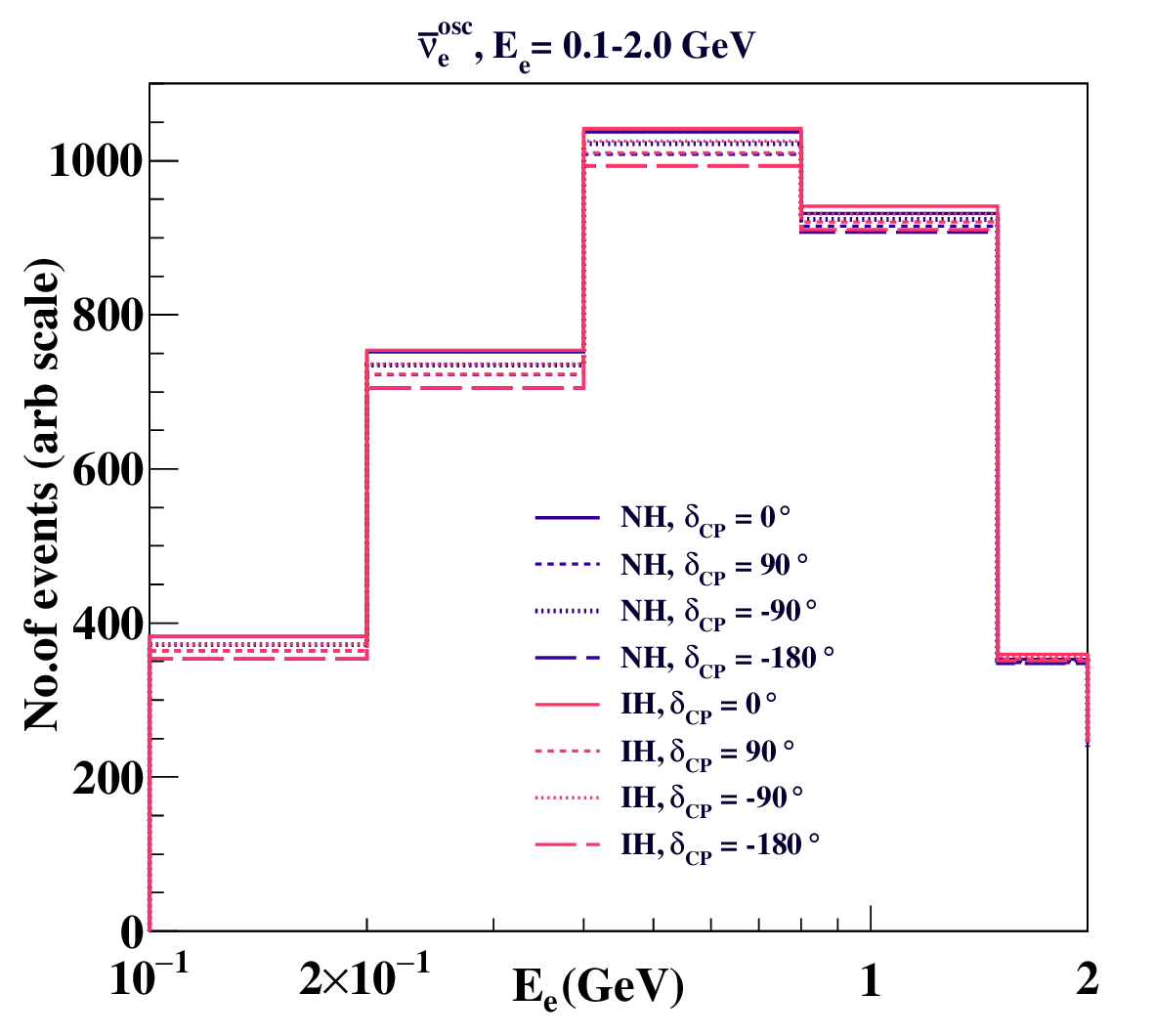}

\includegraphics[width=0.45\textwidth,height=0.445\textwidth]{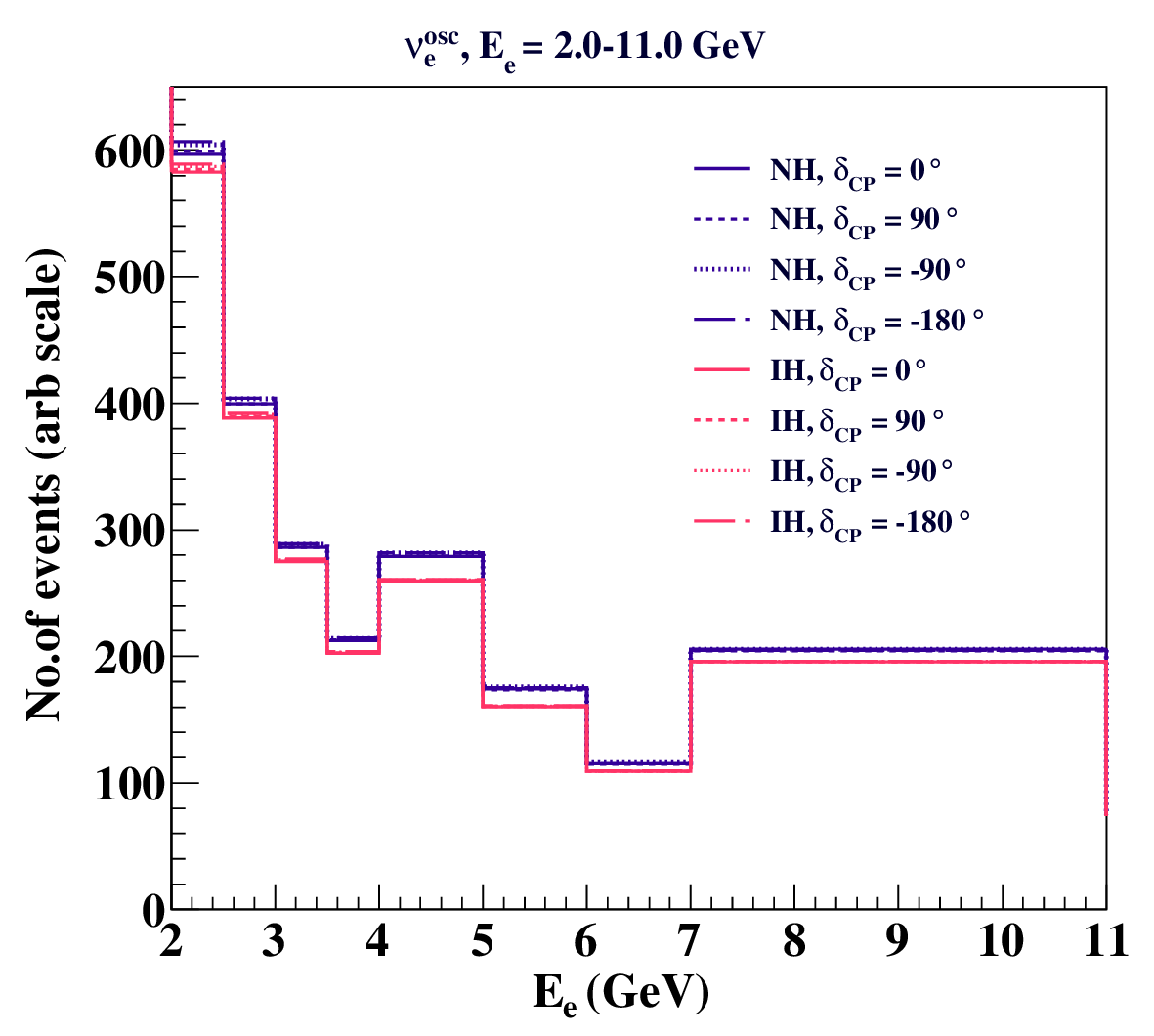}
\includegraphics[width=0.45\textwidth,height=0.445\textwidth]{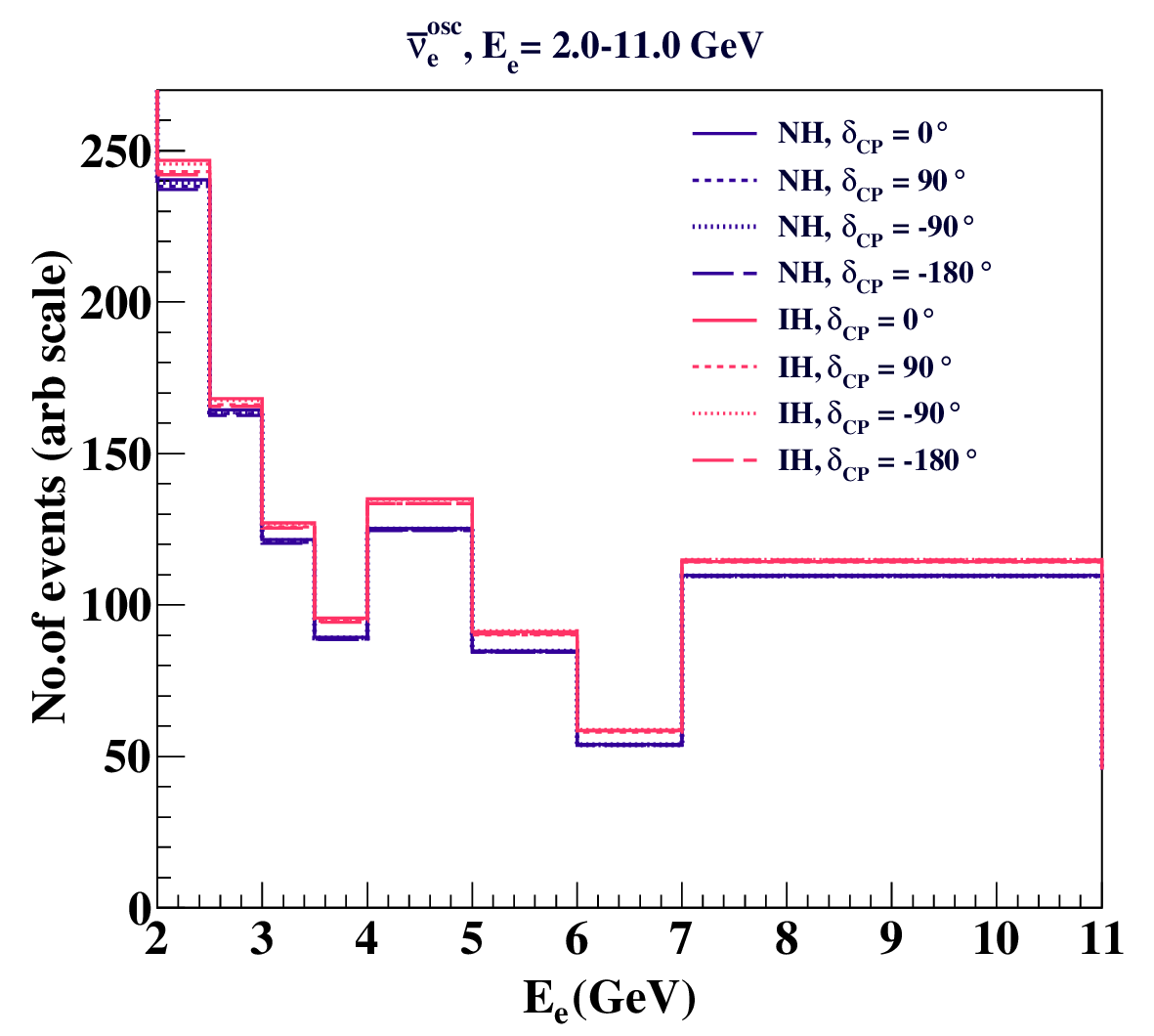}
\caption{Oscillated events with different $\delta_{CP}$ values in the
$E_e$ range (top) 0.1--2.0 GeV and (bottom) 2.0--11.0 GeV. The
blue (pink) histograms are for normal (inverted) hierarchy. The left panels are
for $\nu$ events and the right ones are for $\overline{\nu}$ events. Note
that the y-axes are kept different for visibility.}
\label{delcp-hie-nu-nueb-evt}
\end{figure}

This is illustrated with oscillated $\nu_e$ and $\overline{\nu}_e$
events binned in $E_l$ in the ranges 0.1--2.0 and
2.0--11.0 GeV in Fig.~\ref{delcp-hie-nu-nueb-evt}. Here the events
are averaged over all directions. Only $\nu_e$ and $\overline{\nu}_e$
events are shown in this figure. The effect is similar in $\nu_\mu$
and $\overline{\nu}_\mu$ events, but the $\delta_{CP}$ sensitivity is smaller
in $\nu_\mu$ and $\overline{\nu}_\mu$ events. It can be seen from the figure
that in the lower energy range 0.1--2.0 GeV, the NH and IH spectra with the same
true $\delta_{CP}$ are the same, while the spectra for different true
$\delta_{CP}$ values differ. In the higher energy range from 2.0--11.0 GeV,
this effect is reversed. Thus in the lower energy range we can measure
$\delta_{CP}$ irrespective of hierarchy and in the higher energy range
hierarchy can be determined irrespective of $\delta_{CP}$. This is a
unique signature provided only by atmospheric neutrinos: that neutrinos
of different energies from the same source can be used to probe different
oscillation parameters unambiguously. The flux of atmospheric neutrinos
is smaller than that of the accelerator neutrino experiments; however, the
simultaneous availability of a wide range of energies ($E$) and baselines
($L$) and $\nu$ and $\overline{\nu}$ and of different neutrino flavours,
is a great advantage. In addition to the $\delta_{CP}$ sensitivity from
the accelerator LBL experiments we can add the sensitivities from low
atmospheric neutrino experiments thus increasing the global sensitivity
towards this parameter.

\section{Sensitivity to $\delta_{CP}$ with low energy atmospheric neutrinos}
\label{chi2}

It has been shown in the previous sections that at low energies ($E_\nu<$
1 GeV) the events spectra from different values of $\delta_{CP}$ can be
distinguished from one another \emph{independent} of the neutrino mass
hierarchy. This means that a good sensitivity to $\delta_{CP}$ can be
obtained by analyzing low energy atmospheric neutrino events. We now
proceed to quantify this sensitivity through a simple $\chi^2$ analysis.
No details of detectors are included; the aim is to
establish the $\delta_{CP}$ dependence {\em in principle} through 
an analysis of low energy atmospheric neutrino events.

The events of interest here are those from the charged current (CC) interactions of
$\nu_\mu,\overline{\nu}_\mu,\nu_e$ and $\overline{\nu}_e$. The sensitivity
to $\delta_{CP}$ comes mainly from CC $\nu_e$ and $\overline{\nu}_e$
events. Since the atmospheric neutrino flux contains both $\nu_e$ and
$\nu_\mu$ neutrinos and anti-neutrinos, the $\nu_e$ events detected at
the detector can be from the direct $\nu_e\rightarrow\nu_e$ survived
events as well as from the $\nu_\mu\rightarrow\nu_e$ oscillated events.
The number of charged current (CC) $\nu_e$ events detected is
given by :
\begin{equation}
{\cal {N}}^e =t\times{n_d}\times \int d\sigma_{\nu_e}\times
   \left[P_{ee}\frac{d^2\Phi_e}{dE_\nu~d\cos\theta_\nu}+
   P_{\mu e}\frac{d^2\Phi_\mu} {dE_\nu~d\cos\theta_\nu}\right]~,
\label{toteve}
\end{equation}
where $t$ is the exposure time, $n_d$ is the number of targets in
the detector, $d\sigma_{\nu_e}$ is the differential neutrino
interaction cross section (typically differential in $E_l$,
$\cos\theta_l$, or both), and $d\Phi_{\nu_\mu}$ and $d\Phi_{\nu_e}$ are
the $\nu_\mu$ and $\nu_e$ fluxes. A similar expression holds for muon
neutrino and anti-neutrino events as well.

Events are simulated using the NUANCE \cite{nuance} neutrino generator;
here 5000 kton-years of unoscillated events are generated using Honda
fluxes and a generic isoscalar target, and scaled down to 500 kton years to
reduce fluctuations. ``Data'' is simulated with the central values of
the parameters shown in Table.~\ref{osc-par-3sig} and fitted to the
``theory'' events which are generated by varying the oscillation
parameters in their respective 3$\sigma$ ranges.

\begin{table}[htp]
\centering
\begin{tabular}{|c|c|c|} \hline
   Parameter & True value & Marginalization range \\ \hline
   $\theta_{13}$ & 8.5$^\circ$ & [7.80$^\circ$, 9.11$^\circ$] \\
   $\sin^{2}\theta_{23}$ & 0.5 & [0.39, 0.64] \\ $\Delta{m^2_{eff}}$ &
   $2.4\times10^{-3}~{\rm eV}^2$ & [2.3, 2.6]$\times10^{-3}~{\rm eV}^2$\\
   $\sin^{2}_{12}$ & 0.304 & Not marginalised \\ $\Delta{m^{2}_{21}}$
   & $7.6\times10^{-5}~{\rm eV}^2$ & Not marginalised \\ $\delta_{CP}$
   & 0, $\pm90^\circ$, $\pm180^\circ$ & [-180$^\circ$, 180$^\circ$] \\
   \hline
\end{tabular}
\caption{\small True values and 3$\sigma$ ranges of parameters
used to generate oscillated events. Values except that of
$\delta_{CP}$ are taken as in Ref.~\cite{a3-pre}. For the oscillation
analysis, $\Delta m^2_{31} = \Delta m^2_{\rm eff}+\Delta
m_{21}^2\left(\cos^2\theta_{12}-\cos\delta_{CP}\sin\theta_{13}
\sin2\theta_{12}\tan\theta_{23}\right)~; \Delta m^2_{32}  =  \Delta
m^2_{31}-\Delta m^2_{21}$, for normal hierarchy when $\Delta{m}_{\rm
eff}^2 > 0$. When $\Delta{m}_{\rm eff}^2 < 0$, $\Delta m^2_{31}
\leftrightarrow -\Delta m^2_{32}$ for inverted hierarchy.}
\label{osc-par-3sig}
\end{table}

\begin{figure}[hbp] \centering
\includegraphics[width=0.4\textwidth,height=0.4\textwidth]{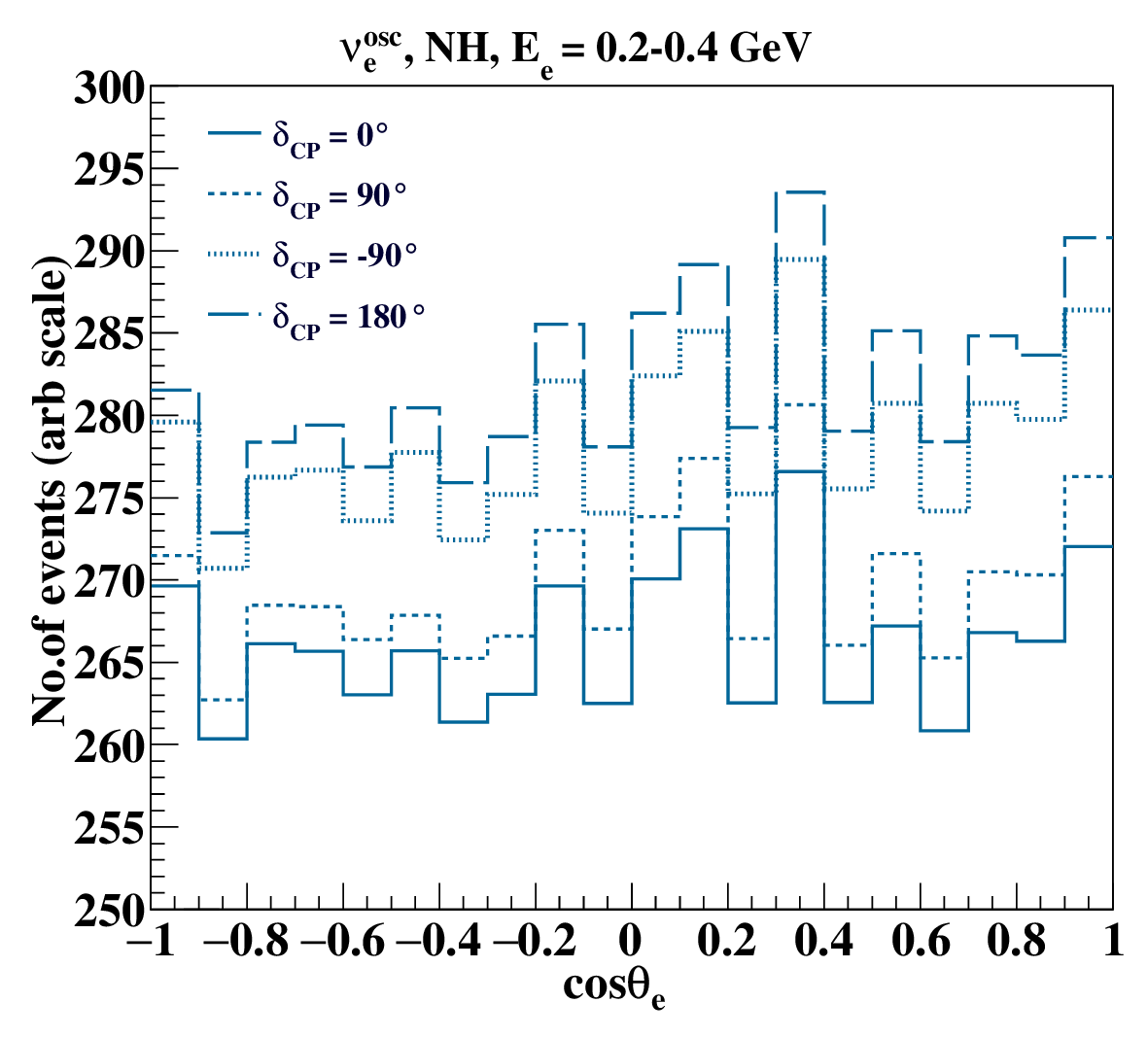}
\includegraphics[width=0.4\textwidth,height=0.4\textwidth]{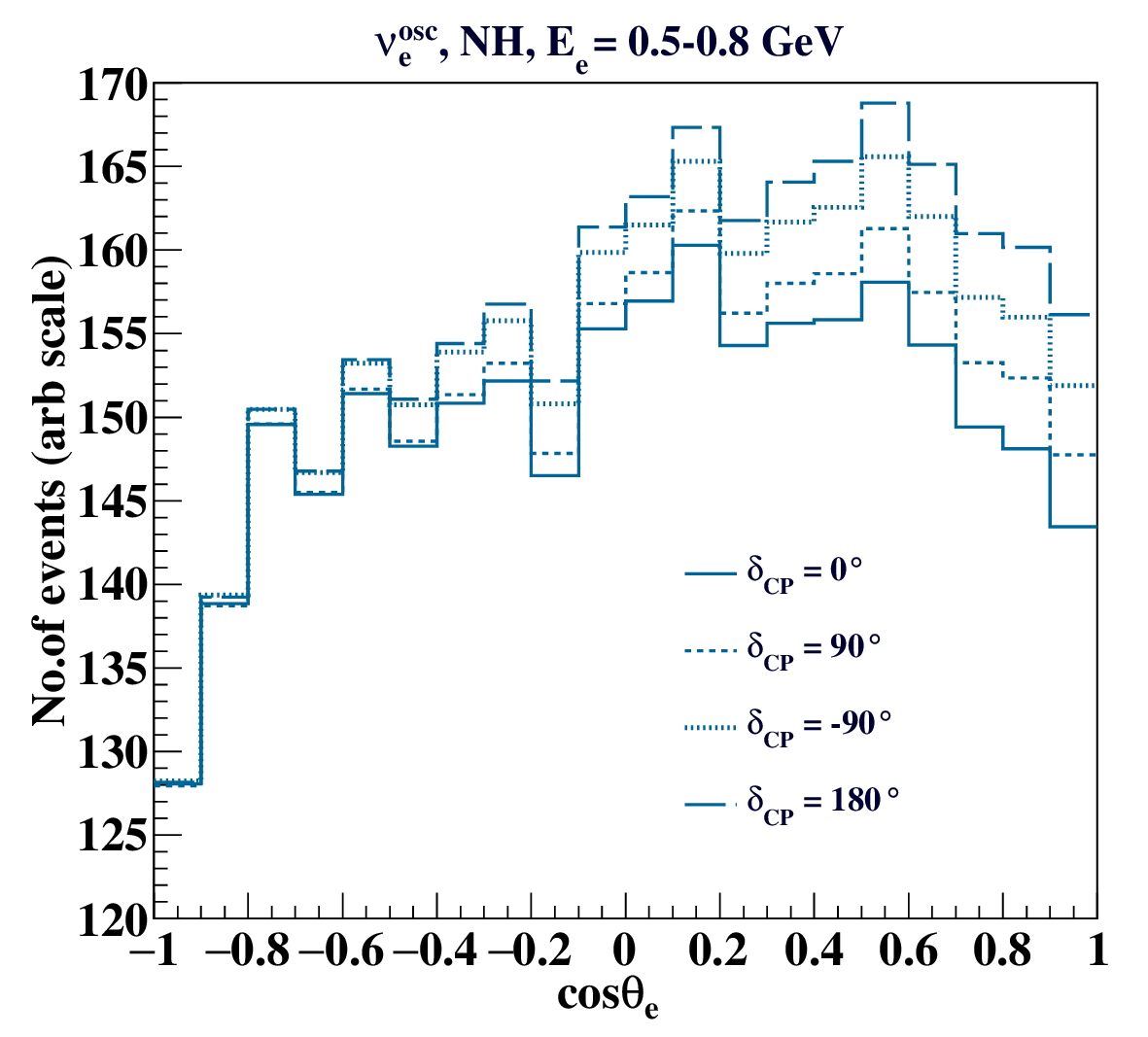}

\includegraphics[width=0.4\textwidth,height=0.4\textwidth]{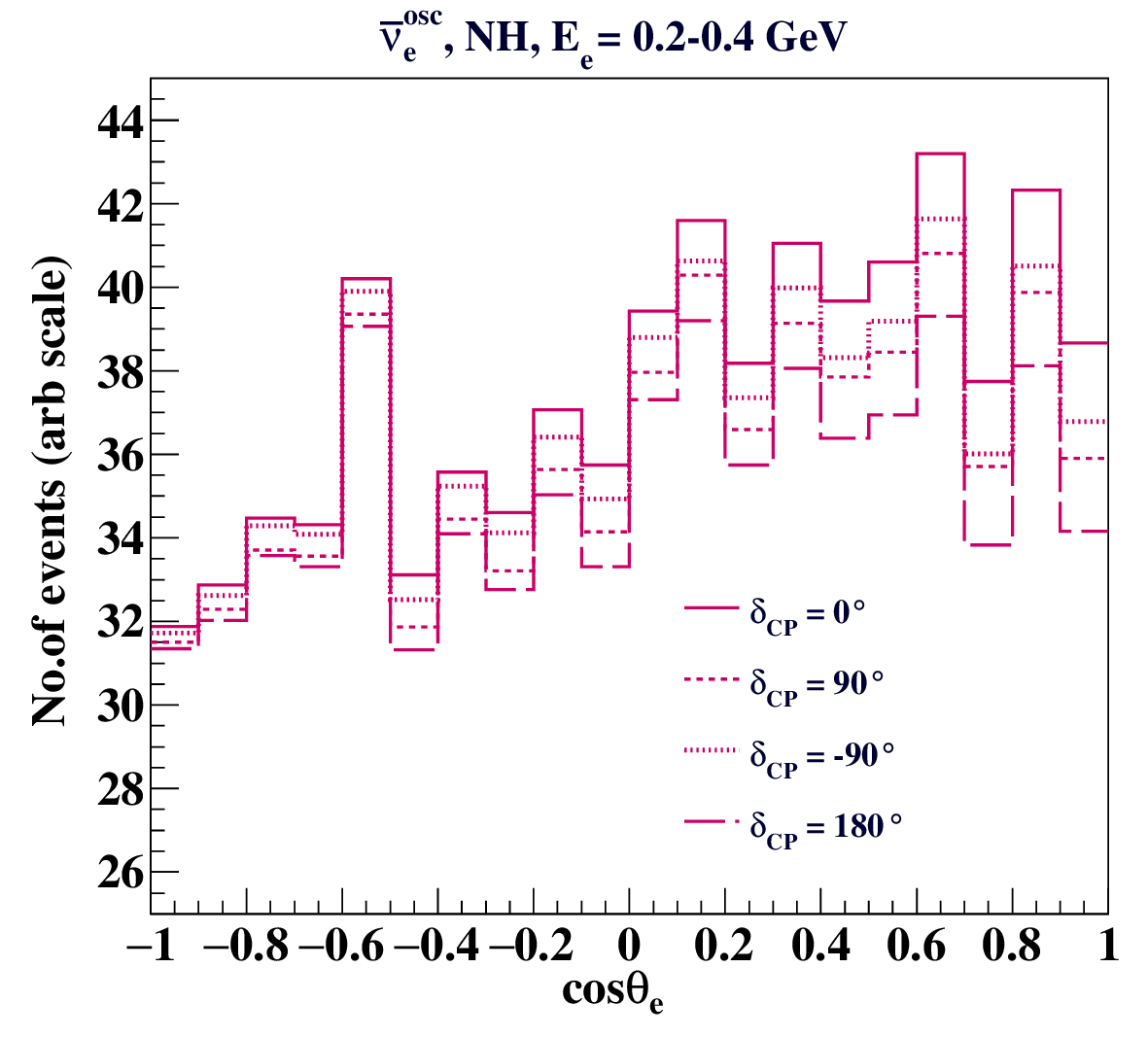}
\includegraphics[width=0.4\textwidth,height=0.4\textwidth]{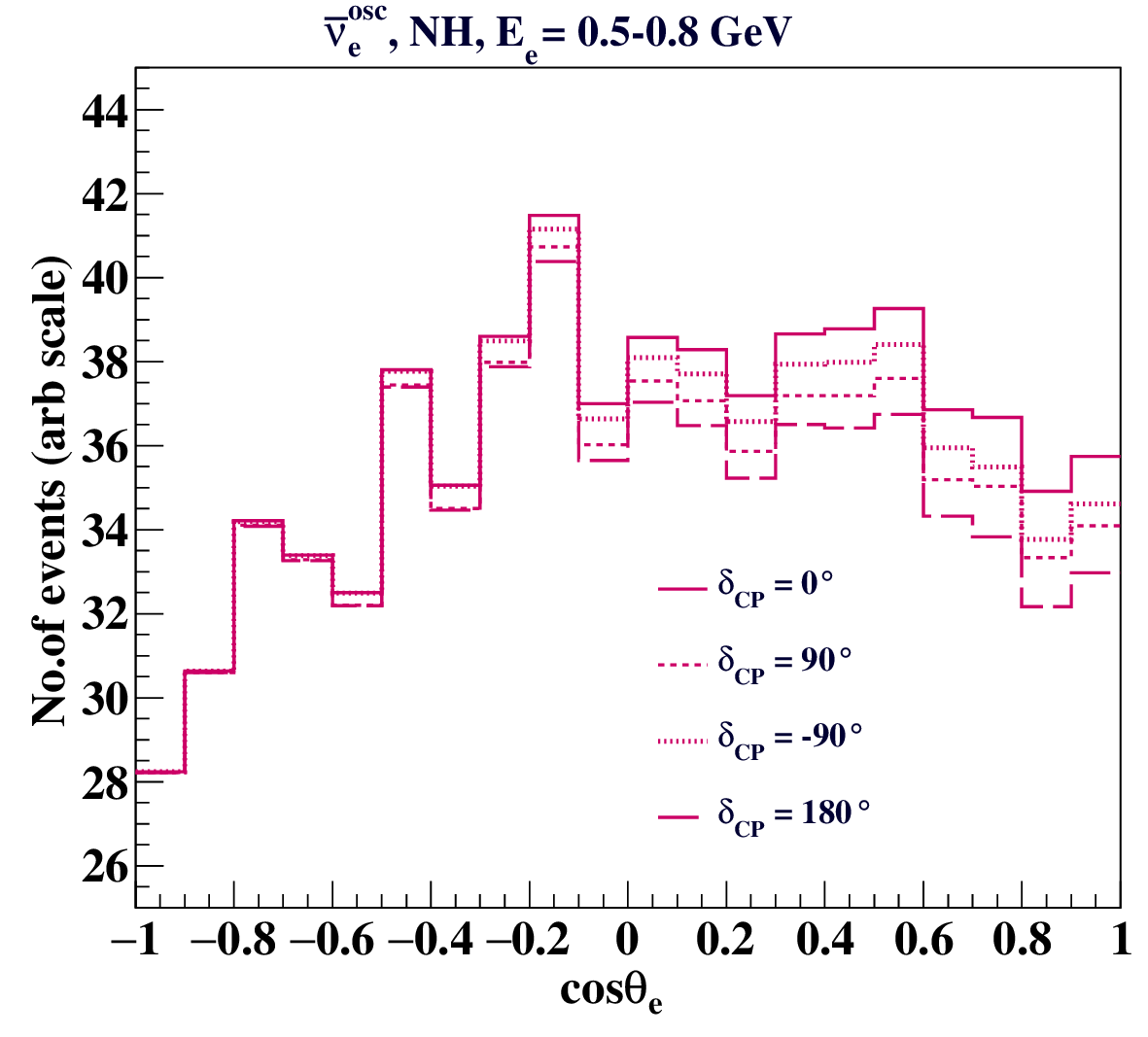}

\caption{Electron events for different true $\delta_{CP}$ values as a
function of $\cos\theta^{obs}_l$ for $E^{obs}_l$ = 0.2--0.4 and 0.5--0.8
GeV; $l=e$. The top row is for $\nu$ events and the bottom one for $\overline{\nu}$.
The y-axes are not the same.}
\label{eve-e}
\end{figure}

\begin{figure}[htp] \centering
\includegraphics[width=0.4\textwidth,height=0.4\textwidth]{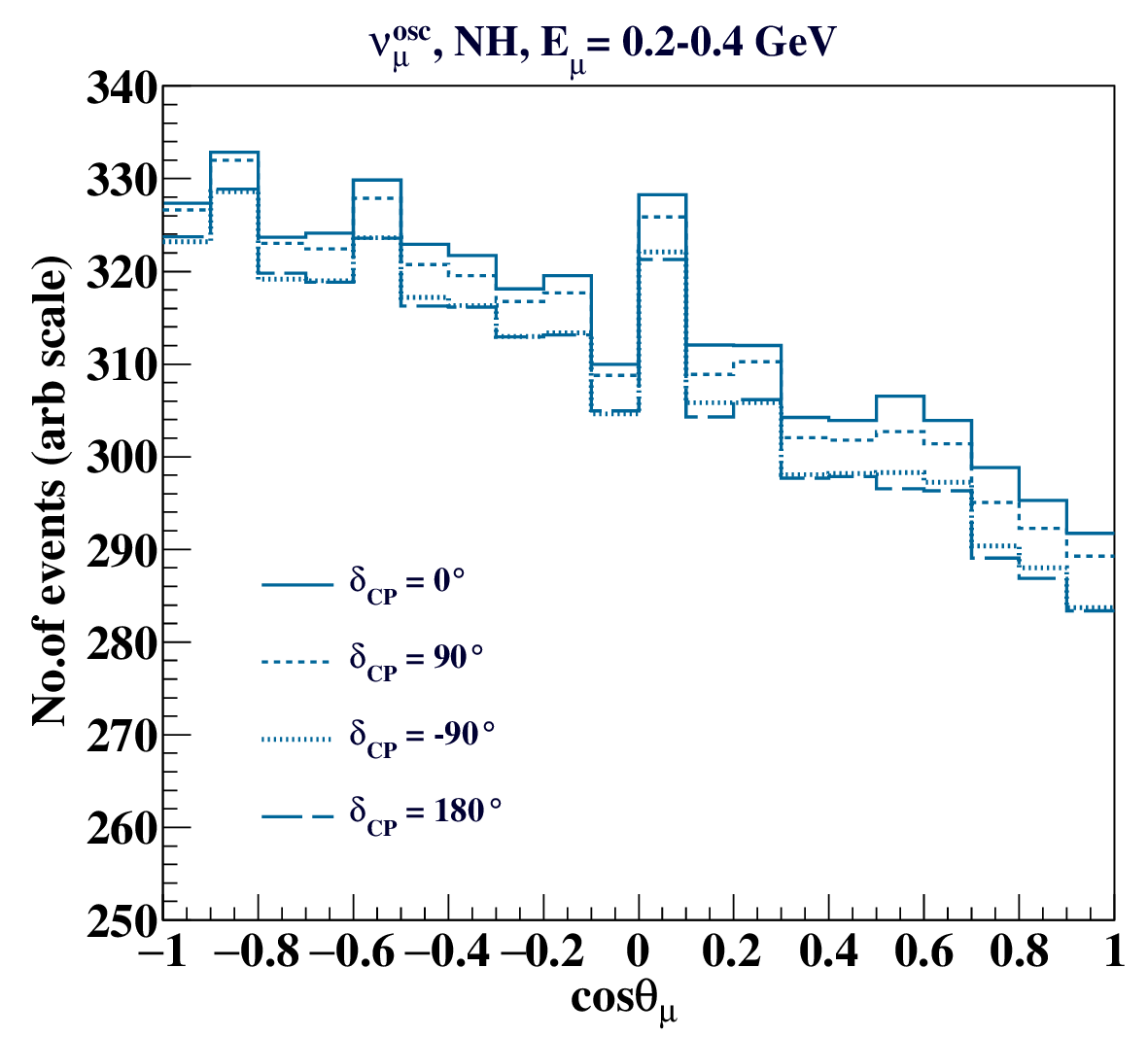}
\includegraphics[width=0.4\textwidth,height=0.4\textwidth]{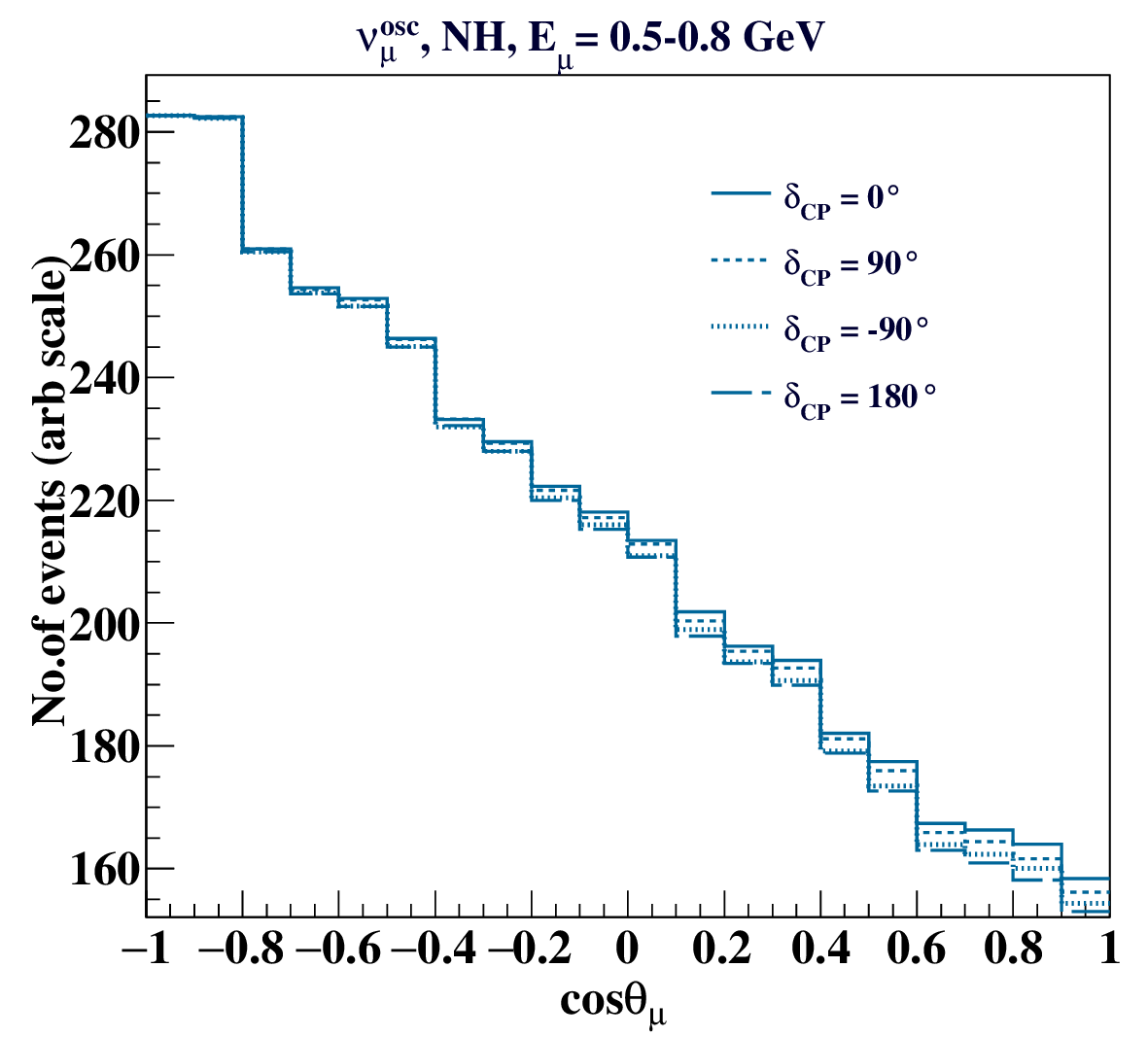}

\includegraphics[width=0.4\textwidth,height=0.4\textwidth]{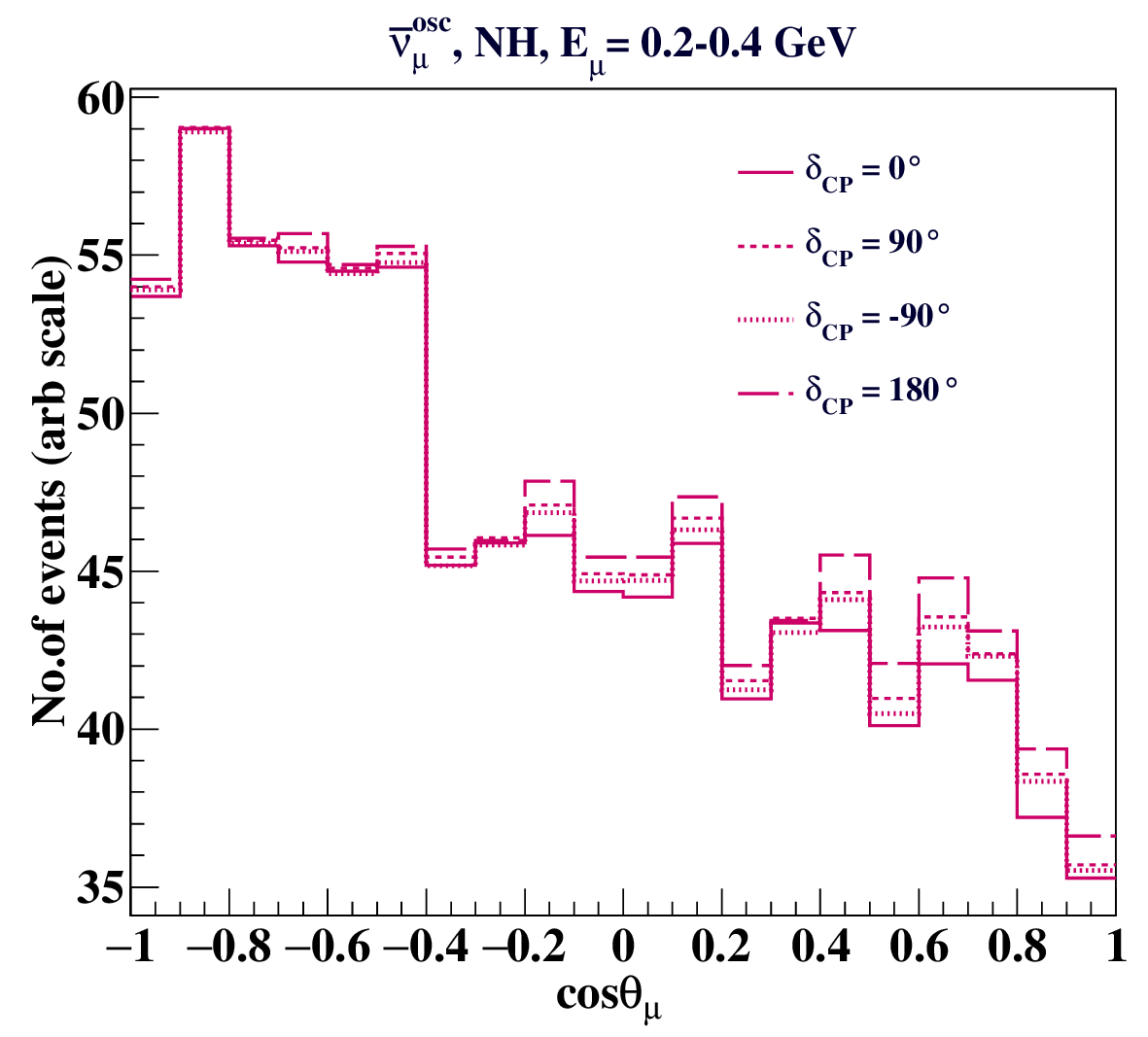}
\includegraphics[width=0.4\textwidth,height=0.4\textwidth]{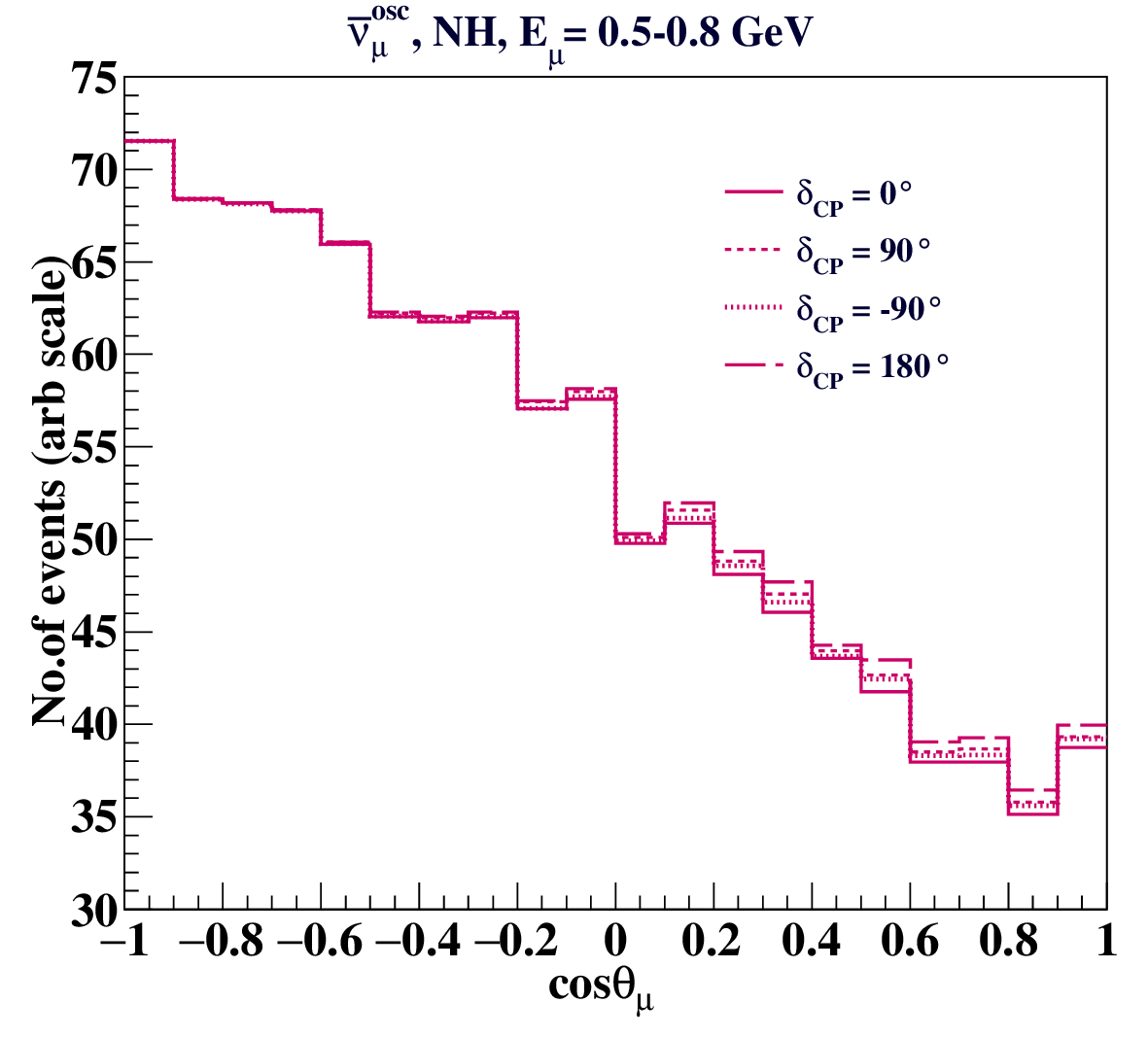}

\caption{As in \protect Fig.~\ref{eve-e}, for muon events, $l=\mu$.
Again, the $y$ axes scales are not the same.
}
\label{eve-mu}
\end{figure}

The event generation and application of oscillations on events
are performed as described in \cite{hi-mu}. The oscillated events
are binned in $(E^{obs}_l,\cos\theta^{obs}_l,E'^{obs}_{had})$, where
$E^{obs}_l,\cos\theta^{obs}_l$ are the energy and direction of the lepton
in the final state, $l=e,\mu$; and $E'^{obs}_{had}$ is the observed
final state hadron energy. The bins used for this analysis are shown
in Table~\ref{binning}. Typical events spectra as a function of
$\cos\theta_l$ for different values of $E_l$, 0.2--0.4 and 0.5--0.8 GeV
respectively, are shown in Figs.~\ref{eve-e} and \ref{eve-mu}.

\begin{table}
\begin{tabular}{|c|c|c|c|} \hline
Observable & Range & Bin width & No.of bins \\ 
\hline
& [0.1, 0.2] & 0.1 & 1 \\ 
& [0.2, 0.4] & 0.2 & 1 \\ 
$E^{obs}_{l}$ (GeV)& [0.4, 0.5] & 0.1 & 1 \\ 
(17 bins) & [0.5, 1.0] & 0.3 & 2 \\ 
& [1, 4] & 0.5 & 6 \\ 
& [4, 7] & 1 & 3 \\ 
& [7, 11] & 4 & 1 \\
& [11, 12.5] & 1.5 & 1 \\ 
& [12.5, 15] & 2.5 & 1 \\ 
& [15, 30] & 15 & 1 \\
\hline 
$\cos\theta^{obs}_{\mu}$ & [-1.0, 1.0] & 0.10 & 20 \\ 
(20 bins) & & & \\ 
\hline & [0, 2] & 1 & 2 \\
$E'^{obs}_{had}$ (GeV) & [2, 4] & 2 & 1\\ 
(4 bins) & [4, 15] & 11 & 1 \\
\hline
\end{tabular}
\caption{The binning scheme used in the analysis. Note that the hadron
energy bins are only relevant for the higher energy sample.}
\label{binning}
\end{table}

It is clearly evident that the effect of $\delta_{CP}$ is more in the
electron type events than in the muon type events. For a given type of
event the separation between the spectra with different $\delta_{CP}$ is
more at lower energies and decreases with the increase of energy. Also
at very low energies the difference is consistent for both the up
and down directions, whereas at higher energies, the effect is more
in the up direction. Because of these consistent differences we can
distinguish different $\delta_{CP}$ values when the events are binned 
in final state lepton direction.

\subsection{$\chi^2$ analysis}

A Poissonian $\chi^2$ analysis assuming an isoscalar detector and no
systematic uncertainties was performed. This is a hypothetical case
to understand how much sensitivity could be obtained under perfect
conditions. The perfect detector has 100\% reconstruction efficiency for
all events and has perfect energy and direction resolutions. In addition to
these, it is assumed that there is perfect separation between CC $\nu_e$,
$\overline{\nu}_e$, $\nu_\mu$ and $\overline{\nu}_\mu$ events. Hence the $\chi^2$
can be expressed as :


\begin{equation}
\chi^2_{l\pm} = \sum_{i}\sum_{j} \sum_{k}
2\left[\left(T^{l\pm}_{ijk}-D^{l\pm}_{ijk}\right)-
D^{l\pm}_{ijk}\ln\left(\frac{T^{l\pm}_{ijk}}{D^{l\pm}_{ijk}}\right)\right]~,
\end{equation}

where $i,j,k$ are the indices corresponding to
$E_l,\cos\theta_l,E^{had}$ bins respectively, the last being the total
hadronic energy in the final state which is not relevant for the low
energy events of interest here. Here $l=e,\mu$ are the final state leptons; 
$T^{l\pm}_{ijk}$ and $D^{l\pm}_{ijk}$  are the theory and ``data'' events 
respectively; $+$ stands for anti-neutrino events and $-$ for neutrino events. 
When we can separate neutrinos from anti-neutrinos, the corresponding $\chi^2$s
can be found out separately as shown here. The total $\chi^2_l$ for $l$ type of 
events is then :
\begin{equation}
\chi^2_{l\delta_{CP}}= \chi^2_{l+}+\chi^2_{l-}~.
\label{chisq-tot}
\end{equation}

\subsection{Results - sensitivity to $\delta_{CP}$}

The sensitivity to the CP phase has been performed with two different
true values $\delta^{true}_{CP}=0,-90^\circ$. Analyses
with all parameters fixed as well as parameters other than $\delta_{CP}$
marginalised in their 3$\sigma$ ranges are also performed. Two different
cases are considered, the first in which $\nu_e,\overline{\nu}_e,\nu_\mu$ and
$\overline{\nu}_\mu$ all can be separately identified. In this case muon charge
identification will help in separating $\nu_\mu$ and $\overline{\nu}_\mu$. In
the second case there is separation between $e$ type and $\mu$ type,
but $\nu_e$ and $\overline{\nu}_e$ cannot be separated from each other and
for muons there is no charge identification to separate $\nu_\mu$ and
$\overline{\nu}_\mu$. A comparison of the $\chi^2$ sensitivities are shown
in the following figures.

\begin{figure} \centering
\includegraphics[width=0.45\textwidth,height=0.45\textwidth]{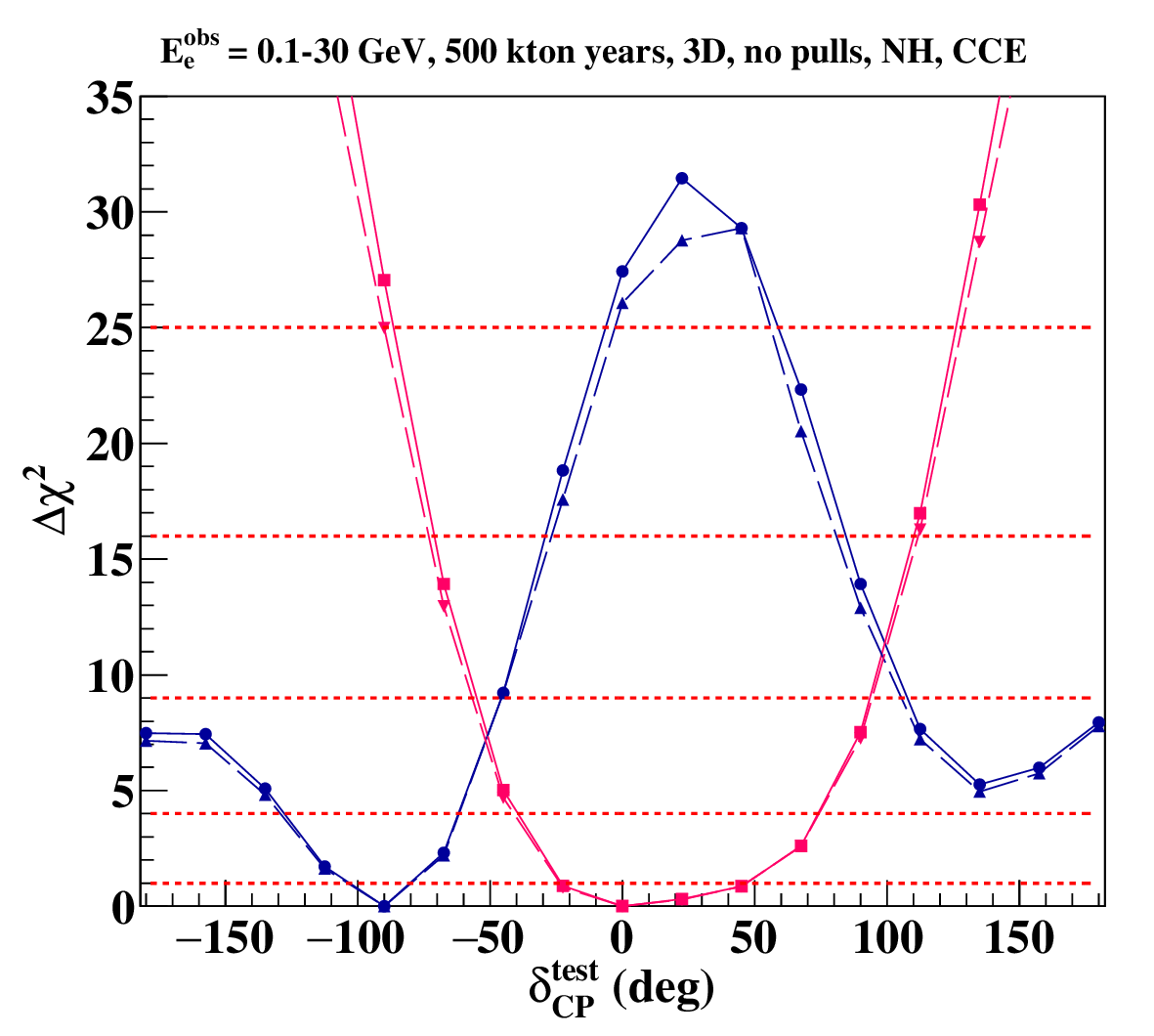}
\includegraphics[width=0.45\textwidth,height=0.45\textwidth]{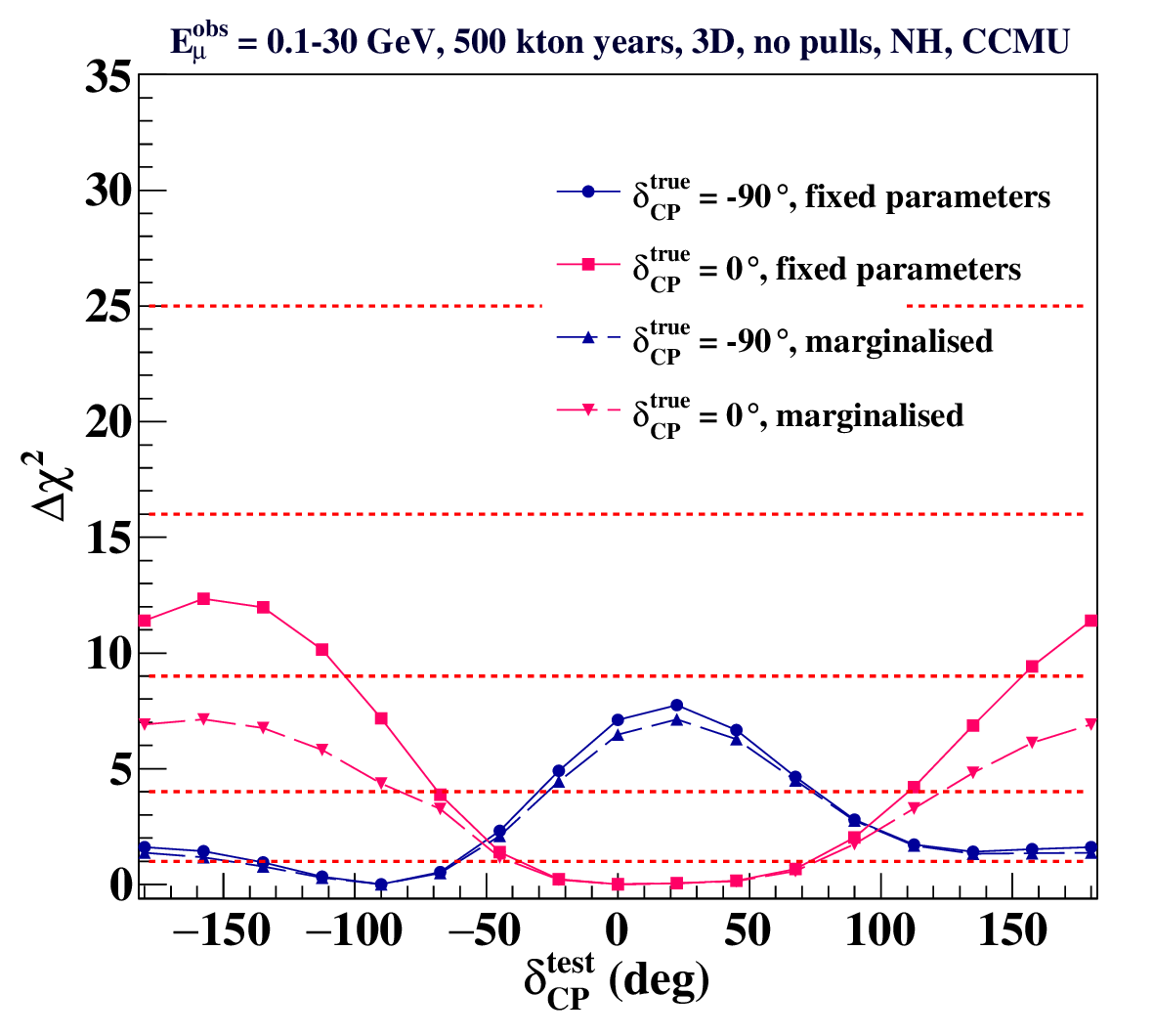}
\caption{\small $\Delta\chi^2$ vs $\delta^{true}_{CP}$ (deg) from
CC $\nu_e+\overline{\nu}_e$ (left) and CC $\nu_\mu+\overline{\nu}_\mu$ (right)
events obtained with 500 kton year of an isoscalar detector;
and with cid, with fixed parameters (solid curves) and marginalisation
(dashed curves). Here $\delta^{true}_{CP}=0,~-90^{\circ}$ (deg). Note that the Y-scales are different.}
\label{chi2-e-mu-wcid}
\end{figure}

The sensitivity to $\delta_{CP}$ when $\nu_e$ ($\nu_\mu$) and
$\overline{\nu}_e$ ($\overline{\nu}_\mu$) can be separated is shown in the left
(right) panel of Fig.~\ref{chi2-e-mu-wcid}. From the figure it can be
readily seen that electron type events have very high sensitivity to
$\delta_{CP}$ as compared to muon type events, which is expected. But at
the same time it is appreciable how muon type events can contribute to
$\delta_{CP}$ sensitivity. This is by virtue of adding the low energy
neutrino events which are sensitive to $\delta_{CP}$. Consider the
situation where we have an atmospheric muon neutrino detector only. If
this detector can be designed in such a way as to detect neutrinos in
energy range 0.1 -- 30.0 GeV, especially those below 1 GeV and can be
magnetised, it can give a very good sensitivity to $\delta_{CP}$ from
muon events alone at low energy, and to the MH at higher energies.

If the true value of $\delta_{CP}$ is $-90^\circ$, then the
parameter space except that from $-135^\circ$ to $-60^\circ$ can be excluded above
2$\sigma$. Whereas electron type events can exclude the same region around
2$\sigma$, the contribution from muon type events is important, since,
when combined with that from electron type events, the $\chi^2$ increases,
thus enabling the exclusions better. 

\begin{figure}
\centering
\includegraphics[width=0.45\textwidth,height=0.45\textwidth]{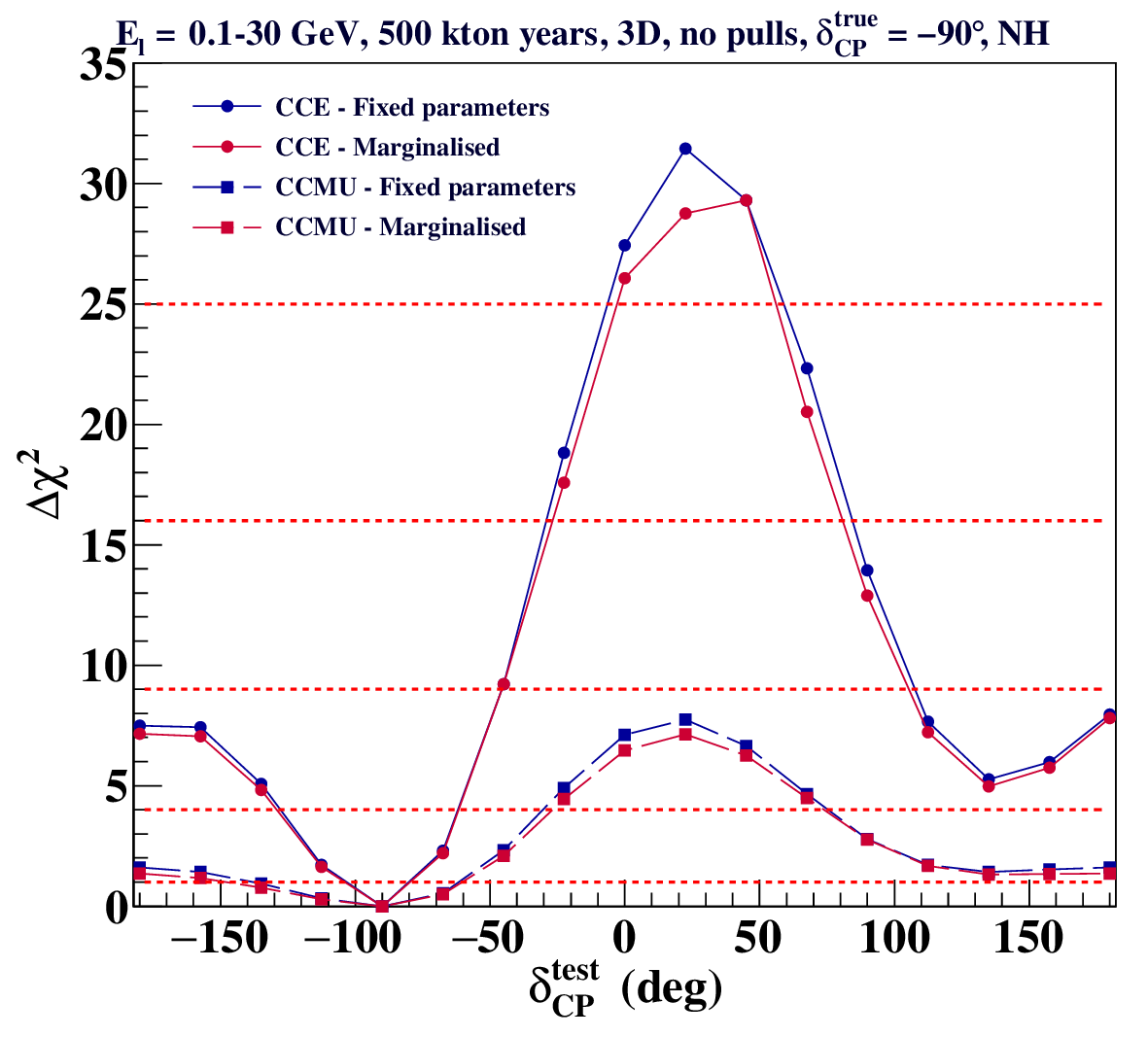}
\includegraphics[width=0.45\textwidth,height=0.45\textwidth]{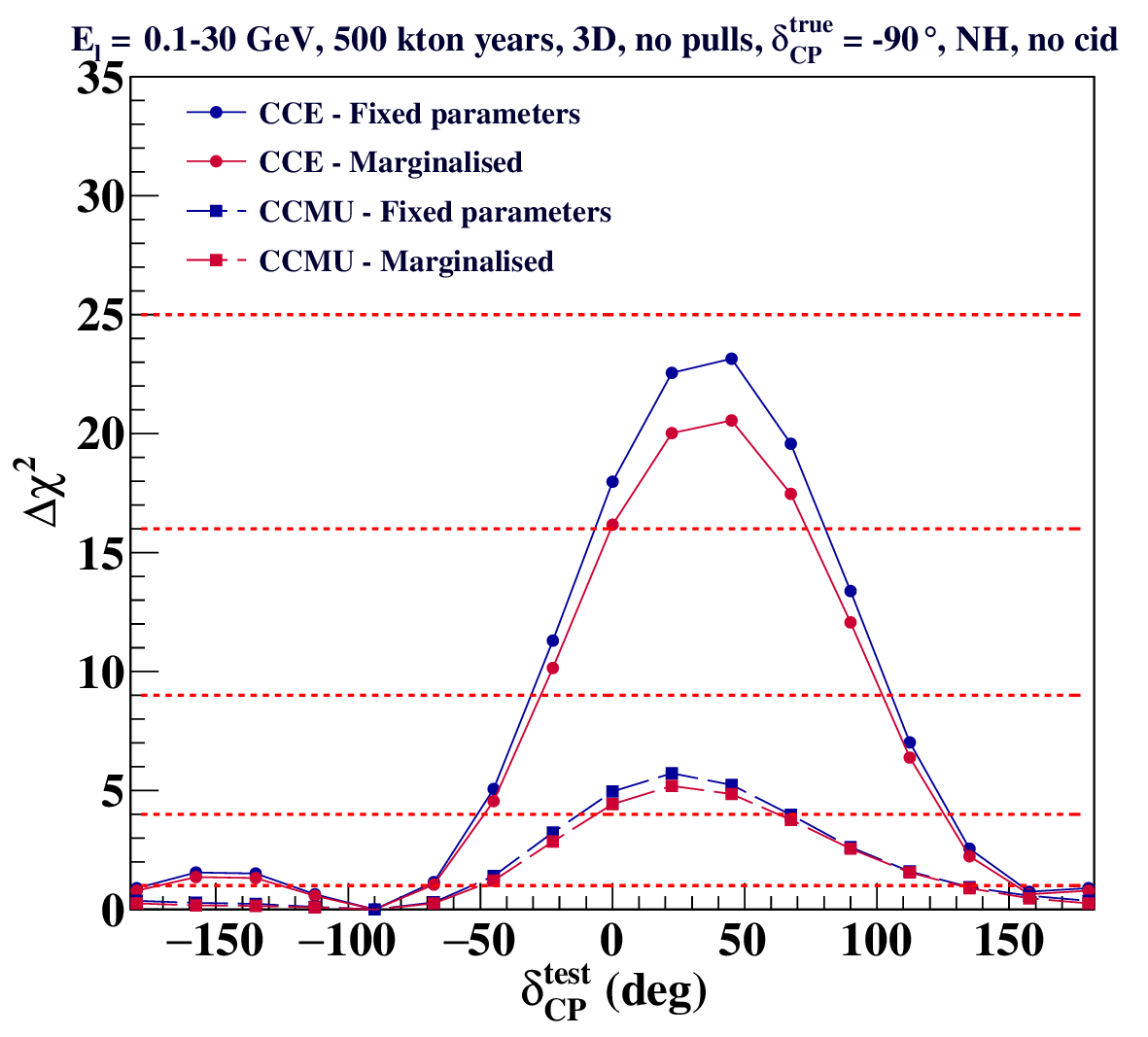}
\caption{$\Delta\chi^2$ vs $\delta^{true}_{CP}$ (deg) from CC
$\nu_e+\overline{\nu}_e$ (solid) and CC $\nu_\mu+\overline{\nu}_\mu$ (dot dashed)
events obtained with 500 kton year of an isoscalar detector
and (left) with cid and (right) no cid. Here $\delta^{true}_{CP}=-90^{\circ}$
(deg) is assumed.}
\label{chi2-e-mu-no-cid}
\end{figure}

The sensitivity to $\delta_{CP}$ when there is separation between
electron type and muon type events but neutrino events cannot be
separated from anti-neutrino events is shown in
Fig.~\ref{chi2-e-mu-no-cid}. Then the sensitivity to $\delta_{CP}$ is
smaller, as compared to the case where $\nu$ and $\overline{\nu}$ can be
identified separately. This is shown in Fig.~\ref{chi2-e-mu-no-cid}
where $\delta^{true}_{CP}=-90^\circ$.  Here it can be seen that the
region $\delta_{CP}\approx[-30^\circ,~80^\circ]$ could be excluded above
4$\sigma$ with $\nu_e-\overline{\nu}_e$ separation, but it reduces to below
4$\sigma$ when they cannot be. A similar result holds for muon type events.
       
From Figs.~\ref{chi2-e-mu-wcid} and \ref{chi2-e-mu-no-cid}, it can
be seen that a detector where we can separate electron type events
from muon type events and also neutrinos and anti-neutrinos in both
cases will give a better sensitivity to $\delta_{CP}$. Of course, this
sensitivity will be reduced when realistic detector resolutions and
systematic errors are taken into account, but the fact to note is the
large discrimination of $\delta_{CP}$ possible with low energy atmospheric 
neutrinos, due to the large $\chi^2$ involved. Also note that 
bin-to-bin correlations among the data will not affect the results since 
the sensitivity to $\delta_{CP}$ is such that the events in
all bins are systematically larger or smaller for a given $\delta_{CP}$.
The separation between $\nu_\mu$ and $\overline{\nu}_\mu$ can be easily 
achieved by having a magnetised detector which will help in identifying the 
charge of the muon such as with the proposed ICAL detector at INO \cite{ical}. 
Compared to muon type events it is difficult to separate $\nu_e$ from
$\overline{\nu}_e$. One of the techniques which can be used to separate
$\nu_e$ and $\overline{\nu}_e$ is to dope the detector material say water
with Gadolinium (Gd) \cite{egads,egads1,egads-old,gadzooks, mori-thesis,
Xu-thesis}. The charged current interaction of $\overline{\nu}_e$ on a free
proton produces a thermal neutron: $\overline{\nu}_e+p\to{e}^++n$, which can
be captured on Gd. The reactions : 
\begin{eqnarray} \nonumber
n+^{155}Gd & \to & ^{156}Gd+\gamma~,\\ \nonumber
n+^{157}Gd & \to & ^{158}Gd+\gamma~,
\end{eqnarray}
produce $\gamma$ rays of energies 8.5 and 7.9 MeV respectively. Then
$\overline{\nu}_e$ can be identified by the coincident detection of $e^+$ and
the $\gamma$. This reaction happens only for $\overline{\nu}_e$
and can be used to separately identify $\nu_e$ and $\overline{\nu}_e$. In
addition, the quasi-elastic cross sections for $\overline{\nu}_e p$ are
proportional to $E_\nu^2$ and are large compared to other processes,
especially at low energy, where the cross section is linearly dependent
on $E_\nu$ and so the events sample is large as well. This technique has
been proposed for detecting supernova neutrinos \cite{mori-thesis,SN-Gd},
but it can be used to detect low energy atmospheric $\overline{\nu}_e$ also
and can be employed in Super-K \cite{SK} and Hyper-K \cite{HK}.

\section{Conclusions}\label{conclusions}

A study of how low energy atmospheric neutrinos can be used to determine
the Dirac CP violating phase $\delta_{CP}$ in the leptonic sector
is performed. It is seen that the events spectra binned according to
the final state lepton direction shows consistent distinction
between various values of $\delta_{CP}$. This allows a precise
determination of $\delta_{CP}$. Also the major issue of hierarchy
ambiguity with $\delta_{CP}$ vanishes at sub GeV energies enabling
a clean measurement of $\delta_{CP}$. For a perfect detector a very
good $\chi^2$ is obtained for $\delta_{CP}$, the major contribution
coming from electron like events. Muon like events also contribute even
though in a less sensitive way. But it is very important to analyse all
possible events since neutrino experiments are low counting ones and every
event adds to the statistics of the experiment. It was also found that
when $\nu_e,\overline{\nu}_e,\nu_\mu$ and $\overline{\nu}_\mu$ can be separated
from one another, $\delta_{CP}$ sensitivity is higher than the case when $\nu$ and
$\overline{\nu}$ cannot be separated from each other. This necessitates having
special detectors which can distinguish between $\nu_e$ and $\overline{\nu}_e$. Gd
doped water Cerenkov detectors are one class of detectors which can
achieve this.

Accelerator long baseline experiments like DUNE themselves can determine
$\delta_{CP}$ with very high precision because of the large statistics and
the excellent resolutions of the detector. But the sub GeV energy atmospheric 
neutrinos should not be abandoned since they provide an alternate method of 
determining $\delta_{CP}$, {\em independent of the neutrino mass hierarchy} 
and the results will add to the global sensitivity to $\delta_{CP}$ thus 
increasing the overall sensitivity to the parameter. It will be interesting 
to study how well can different atmospheric neutrino detectors probe $\delta_{CP}$
especially in the sub GeV range and whether any modification to detector
configurations will improve the sensitivity to this parameter. A
detailed analysis including detector resolutions and systematic effects
is beyond the scope of this paper and is work in progress.

\section{Acknowledgements}
     We are grateful to Prof.G.~Rajasekaran and Prof.~Rahul Sinha, IMSc
     Chennai for many discussions. We also thank the Journal Club at IMSc
     where this idea was first discussed, and Prof.T.~Kajita for helpful
     comments during the EILH workshop at Aligarh Muslim Unversity. LSM
     thanks Prof.~Jim Libby, IIT Madras, Chennai. She also acknowledges
     Nandadevi cluster which is a part of the compupting facility at IMSc
     Chennai, with which the simulations were performed.

\appendix
\section{Details of hierarchy independence at low energies}\label{appendix:A}

The transition probability in vacuum can be expressed as : 
\begin{eqnarray}
 P^{vac}_{\alpha\beta}&=&-4Re[U_{\alpha2}U^*_{\beta2}U^*_{\alpha1}U_{\beta1}]\sin^2(1.27\Delta{m^2_{21}L/E})\\  
 &&
 -2Re[U_{\alpha3}U^*_{\beta3}(\delta_{\alpha\beta}-U^*_{\alpha3}U_{\beta3})] \\
 && + 2Im[U_{\alpha2}U^*_{\beta2}U^*_{\alpha1}U_{\beta1}]\sin(2.53\Delta{m^2_{21}}L/E).
\end{eqnarray}

Since the probability is independent of $\Delta{m^2_{32}}$, there is no hierarchy ambiguity. 
\begin{eqnarray*}
P_{e\mu}&=&A+B\cos\delta-C\sin\delta=\overline{P}_{\mu e};\\
P_{\mu e}&=&A+B\cos\delta+C\sin\delta=\overline{P}_{e\mu}, 
\end{eqnarray*} where 

\begin{eqnarray*}
 A=c^2_{13}\sin^2(2\theta_{12})(c^2_{23}-(s_{23}s_{13})^2)\sin^2(\delta_{21}/2)+\frac{1}{2}s^2_{23}\sin^2(2\theta_{13}),\\ \nonumber
 B=(1/4)c_{13}\sin(4\theta_{12})\sin(2\theta_{13})\sin(2\theta_{23})\sin^2(\delta_{21}/2),\\ \nonumber
 C=(1/4)c_{13}\sin(2\theta_{12})\sin(2\theta_{13})\sin(2\theta_{23})\sin(\delta_{21}), \\ \nonumber
 \delta_{21}=2.534\Delta{m^2_{21}L/E}.
 \end{eqnarray*}

 $A,B,C$ are only limited only by precisions measurements of oscillation parameters. 
The CP asymmetry can be expressed as : 
\begin{eqnarray}
A_{CP}=\frac{P_{e\mu}-P_{\mu e}}{P_{e\mu}+P_{\mu e}} = -\frac{C}{A+B\cos\delta}\sin\delta \\  
\overline{A}_{CP} = \frac{\overline{P}_{e\mu}-\overline{P}_{\mu e}}{\overline{P}_{e\mu}+\overline{P}_{\mu e}} = \frac{C}{A+B\cos\delta}\sin\delta
\end{eqnarray}
for $\nu$ and $\overline{\nu}$ respectively.

\end{document}